\documentclass[]{aastex631}
\newcommand\red[1]{#1}

\usepackage{mathtools}
\usepackage[caption=false]{subfig}

\newcommand\injdet{$P(det|S_{i,det})$}
\newcommand\injt{$P(det|S_{T})$}
\newcommand{\err}{$P(S_{i,det}|S_{T},\sigma_{d})$}

\begin{document}
\title{Constraining the selection corrected luminosity function and total pulse count for radio transients}

\correspondingauthor{Fengqiu Adam Dong}
\email{fengqiu.dong@gmail.com}

\author[0000-0003-4098-5222]{Fengqiu Adam Dong}
\affiliation{Department of Physics and Astronomy, University of British Columbia, 6224 Agricultural Road, Vancouver, BC V6T 1Z1 Canada}
  
\author[0000-0002-3654-4662]{Antonio Herrera-Martin}
\affiliation{David A.~Dunlap Department of Astronomy \& Astrophysics, University of Toronto, 50 St.~George Street, Toronto, ON M5S 3H4, Canada}
\affiliation{Department of Statistical Science, University of Toronto, Ontario Power Building, 700 University Avenue, 9th Floor, Toronto, ON, Canada M5G 1Z5}

\author{Ingrid Stairs}
\affiliation{Department of Physics and Astronomy, University of British Columbia, 6224 Agricultural Road, Vancouver, BC V6T 1Z1 Canada}

\author[0000-0002-1348-8063]{Radu V. Craiu}
\affiliation{Department of Statistical Science, University of Toronto, Ontario Power Building, 700 University Avenue, 9th Floor, Toronto, ON, Canada M5G 1Z5}
 
\author[0000-0002-1529-5169]{Kathryn Crowter}
\affiliation{Department of Physics and Astronomy, University of British Columbia, 6224 Agricultural Road, Vancouver, BC V6T 1Z1 Canada} 

\author[0000-0003-3734-8177]{Gwendolyn M. Eadie}
\affiliation{David A.~Dunlap Department of Astronomy \& Astrophysics, University of Toronto, 50 St.~George Street, Toronto, ON M5S 3H4, Canada}
\affiliation{Department of Statistical Science, University of Toronto, Ontario Power Building, 700 University Avenue, 9th Floor, Toronto, ON, Canada M5G 1Z5}

\author[0000-0001-8384-5049]{Emmanuel Fonseca}
\affiliation{Department of Physics and Astronomy, West Virginia University, Morgantown, WV 26506-6315, USA}
\affiliation{Center for Gravitational Waves and Cosmology, Chestnut Ridge Research Building, Morgantown, WV 26505, USA}

\author[0000-0003-1884-348X]{Deborah Good}
\affiliation{Physics Department, University of Montana, Charles H Clapp, Building Rm 126, 32 Campus Drive, Missoula, MT 59812 }

\author[0000-0002-2885-8485]{James W. Mckee}
\affiliation{E.A. Milne Centre for Astrophysics, University of Hull, Cottingham Road, Hull, HU6 7RX, UK}
\affiliation{Centre of Excellence for Data Science, Artificial Intelligence and Modelling, University of Hull, Cottingham Road, Hull, HU6 7RX, UK}

\author[0000-0001-8845-1225]{Bradley W. Meyers}
\affiliation{International Centre for Radio Astronomy Research, Curtin University, Bentley, WA 6102, Australia}
\author[0000-0002-8912-0732]{Aaron B.~Pearlman}
\affiliation{Department of Physics, McGill University, 3600 rue University, Montr\'eal, QC H3A 2T8, Canada}
\affiliation{Trottier Space Institute, McGill University, 3550 rue University, Montr\'eal, QC H3A 2A7, Canada}

\author[0000-0002-9761-4353]{David C. Stenning}
\affiliation{Department of Statistics and Actuarial Science, Simon Fraser University, 8888 University Drive, Burnaby, B.C., V5A 1S6 Canada}



\begin{abstract}
Studying transient phenomena, such as individual pulses from pulsars, has garnered considerable attention in the era of astronomical big data. Of specific interest to this study are Rotating Radio Transients (RRATs), nulling, and intermittent pulsars. This study introduces a new algorithm named LuNfit, tailored to correct the selection biases originating from the telescope and detection pipelines. Ultimately LuNfit estimates the intrinsic luminosity distribution and nulling fraction of the single pulses emitted by pulsars. \red{LuNfit relies on Bayesian nested sampling so that the parameter space can be fully explored. Bayesian nested sampling also provides the additional benefit of simplifying model comparisons through the Bayes ratio.} The robustness of LuNfit is shown through simulations and applying LuNfit onto pulsars with known nulling fractions. LuNfit is then applied to three RRATs, J0012+5431, J1538+1523, and J2355+1523, extracting their intrinsic luminosity distribution and burst rates. We find that their nulling fraction is 0.4(2), 0.749(5) and 0.995(2) respectively. We further find that a log-normal distribution likely describes the single pulse luminosity distribution of J0012+5431 and J1538+1523, while the Bayes ratio for J2355+1523 slightly favors an exponential distribution. We show the conventional method of correcting selection effects by ``scaling up'' the missed fraction of radio transients can be unreliable when the mean luminosity of the source is faint relative to the telescope sensitivity. Finally, we discuss the limitations of the current implementation of LuNfit while also delving into potential enhancements that would enable LuNfit to be applied to sources with complex pulse morphologies.
\end{abstract}



\section{Introduction} \label{sec:intro}
Neutron stars result from the remnants of massive stars after they have undergone core-collapse supernova. Rapid rotation of neutron stars likely emerges due to angular momentum conservation following stellar collapse. The rapid rotation combined with beamed radiation creates a lighthouse effect and results in pulsed signals at observatories on Earth. The observed emission exhibits a frequency-dependent delay due to cold-plasma dispersion, imparted onto the radio signal as it traverses the interstellar medium. The dispersion is dependent on the integrated line-of-sight electron density and is inversely proportional to the square of the observing frequency (i.e. $\Delta t=\frac{1}{2.41\times 10^{-4}}~\text{s}~\text{DM}$~pc/cm$^{3}~((\nu_1/\text{GHz})^{2}-(\nu_2/\text{GHz})^{2})$ \citep{2004hpa..book.....L} where $\Delta t$ is the time delay and $\nu_1$ and $\nu_2$ are the upper and lower observing frequencies. DM is the dispersion measure). The conventional technique to detect pulsars is to first correct for the dispersion and then fold the observation on the pulsar period to build signal-to-noise (S/N). Therefore, pulsar detection sensitivity can be significantly boosted by longer integration times. Pulsar survey strategies \red{must strike} a balance between integration time and sky coverage when using conventional single-dish radio telescopes \citep{10.1046/j.1365-8711.2001.04751.x,10.1086/498335,10.1088/0004-637X/791/1/67}. Often, but not always, this leads to surveys focussing on the Galactic plane. A counter-example is the Green Bank North Celestial Cap survey which surveyed the whole northern hemisphere \citep{10.1088/0004-637X/791/1/67}. Because of the large sky coverage, \red{the survey only dwelled for} 120~s per pointing. \red{The contemporary pulsar astronomy landscape operates in a distinctly different parameter space} such as the Canadian Hydrogen Intensity Mapping Experiment (CHIME) and the Low-Frequency Array (LOFAR) \citep{10.3847/1538-4357/aad188,10.1051/0004-6361/201935609}. Their wide field of view allows for simultaneous significant sky coverage and long dwell times. With these new capabilities, intermittent pulsars, nulling pulsars, and RRATs become easier to find and characterize.

It has long been known that some pulsars do not emit on every rotation \citep{10.1038/228042a0}. This phenomenon is called nulling, and up until recently, the population of known nulling pulsars has been small compared to the total pulsar population, likely due to the aforementioned survey methods \citep{10.1007/s12036-019-9608-z}. With access to instruments with wide fields of view, this may be changing. For example, \cite{10.3847/1538-4357/abb94f} showed that even long-known pulsars thought to be persistent emitters could exhibit nulling behavior. It is important to understand the nulling phenomenon as this can lead to better models for pulsar emission mechanisms \citep{arXiv:astro-ph/0208344}. On the extreme end of the nulling pulsar spectrum are RRATs, highly intermittent pulsars that can only be detected via their single pulses. As a large portion of the pulsar population likely consists of RRATs \citep{10.1038/nature04440,10.48550/arXiv.1109.6896}, understanding their intrinsic characteristics is vital to understanding the pulsar population. However, RRATs can not be folded to increase S/N and thus may require long observations to detect appreciable amounts of single pulses.

A critical aspect of population analysis lacking in the current literature is probing the intrinsic luminosity function and burst rate of highly intermittent sources using only the detected pulsar single pulses. Many studies of RRATs only describe an observed burst rate \citep[e.g.][]{10.1088/0004-637X/809/1/67,10.3847/0004-637X/821/1/10,10.3847/1538-4357/ac1da6,10.1093/mnras/stad2012} rather than correcting for selection effects to probe the intrinsic values. There have been attempts to probe the intrinsic parameters for nulling pulsars \citep{10.1093/mnras/176.2.249,10.3847/1538-4357/aaab62,10.3847/1538-4357/acbb68} and also another similar class of astrophysical phenomena, repeating Fast Radio Bursts, luminous extragalactic radio bursts with similar burst characteristics to RRATs \citep{10.1038/s41586-021-03878-5}. \red{The most established method described in \cite{10.1093/mnras/176.2.249} relies on forming histogram bins and calculating $\Delta=\rm{ON} - NF\times \rm{OFF}$ for each histogram bin, where $\rm{ON}$ and $\rm{OFF}$ are the histogram bins of the integrated intensities during the on-pulse window and off-pulse window respectively and $NF$ is the nulling fraction. One would then minimize $|\Delta|$ by changing $NF$ to find the preferred nulling fraction.} However, this method and the Gaussian mixture model in \cite{10.3847/1538-4357/aaab62} operate on the folded time series of the pulsar observations and necessarily requires two conditions: the first of these conditions is a period and a correctly identified pulse arrival phase. This is not always possible with RRATs due to their extreme sporadicity \citep[i.e.][]{10.1093/mnras/stad2012}. The second requirement is recording the full-time series for the observation, even during times of no emission. This is problematic for some contemporary commensal instruments such as CHIME/FRB \citep{10.3847/1538-4357/aad188}, which only save data segments around detected single pulses. While the method described in \cite{10.1038/s41586-021-03878-5} looks at only single pulses, they account for the missed pulses by ``scaling up'' by the measure detection fraction. However, we show in section \ref{sec:comparison} that this method can be unreliable, especially when the source has emitted many bursts below the observing telescope detection threshold.
In an attempt to avoid the limitations of earlier works, we implement a new method, \textit{Luminosity and N fit} (LuNfit), that utilizes Bayesian inference using the nested sampling algorithm \citep{https://ui.adsabs.harvard.edu/abs/2004AIPC..735..395S} to constrain the intrinsic single pulse luminosity distribution and the burst rates simultaneously. LuNfit is designed to remove any systematic biases in the single pulse luminosity distribution and burst rate by empirically measuring the selection effects of the telescope and detection pipeline. 

This study is presented in the following way. Section \ref{sec:methods} describes the CHIME instrument and the implementation of LuNfit. We discuss the likelihood and derivation of LuNfit. Section \ref{sec:lum_dist} described the underlying luminosity distributions used. \red{ Section \ref{sec:priors} details the priors used in LuNfit. Section \ref{sec:discrete_N} addresses the issue of discrete parameters. Section \ref{sec:detections} details the CHIME/Pulsar backend, the detection of pulses, and how single pulse characteristics are retrieved following the detection of pulses. Section \ref{sec:injections} details the process of injecting simulated pulses into the CHIME/Pulsar detection pipeline to find the selection biases of the telescope. Section \ref{sec:simulations} details the simulations we performed to show the efficacy and robustness of LuNfit. Section \ref{sec:validation} compares LuNfit results with the known nulling fractions of two pulsars, B1905+39 and J2044+4614. Section \ref{sec:conversion} describes the conversion to flux units.} Section \ref{sec:results} applies LuNfit to three RRATs, J0012+54, J1528+2345, and J2355+1523. We present their intrinsic luminosity distributions and burst rates. Finally, Section \ref{sec:discussion} compares LuNfit and conventional methods of correcting selection biases and discusses LuNfit's potential future applications to pulsars, RRATs, and FRBs.

\section{Methods}\label{sec:methods}
\begin{table}[]
\centering
\caption{}
\label{tab:CHIME-props}
\begin{tabular}{l|l}\hline\hline
                                   & CHIME/Pulsar \\\hline\hline
Receiver noise temperature         & $\sim$50K  \\
Frequency Range                    & 400-800MHz \\
Number of beams                    & 10 (Tracking) \\
Beam width (FWHM)                  & \red{30'(400MHz)-15'(800MHz)}\\
Time resolution                    & 327.68$\mu$s \\
Search Frequency Resolution        & 390.625kHz \\
Coherent Dedispersion              & Yes       \\
\end{tabular}
\end{table}
CHIME is a commensal transit radio telescope situated in British Columbia, Canada. It comprises three backend instruments: CHIME/cosmology, CHIME/FRB, and CHIME/Pulsar \citep{,10.3847/1538-4357/aad188,10.3847/1538-4365/abfdcb,10.3847/1538-4365/ac6fd9}. Our study only utilizes the CHIME/Pulsar backend, which processes up to 10 simultaneous steerable tracking beams as sources transit the CHIME sky. The properties of CHIME/Pulsar are provided in Table \ref{tab:CHIME-props}. The transit nature of the CHIME telescope allows it to cover the whole northern sky every day, and while each observation is only around $\sim 10$~minutes CHIME/Pulsar can achieve daily cadence on astrophysical sources. Every day CHIME/Pulsar can observe $\sim 500$ individual sources. For this study, CHIME/Pulsar collects high-time resolution spectra in the form of Sigproc\footnote{https://sigproc.sourceforge.net/} style filterbank data. CHIME's sensitivity changes relative to the zenith angle of the source; therefore, to maintain consistency within each observation, we use only the data in which pulses from the nominal pulsar position would be visible through the entire frequency range.

For this study, we are focussed on RRATs, which are only detectable via single pulses. \red{A Bernoulli process is sequence of binary random variables i.e. they take only values of 0 or 1. Each pulse from an RRAT can be described as one of these binary variables where 0 represents non-detection, and 1 represents detection. While RRAT pulses have not been modelled as Bernoulli process previously, RRATs are often assumed to follow a Poisson distribution, the extreme case for the Bernoulli process \citep[e.g.][]{10.3847/1538-4357/ac75d1,10.48550/arXiv.2302.12661}. During any observation of a RRAT, any radio telescope will detect only a portion of the emitted pulses due to limited sensitivity. This is called the ``selection' of the telescope. Therefore,} Assuming the selection effects of the telescope are understood, we forward model the intrinsic luminosity function of the RRAT. For N pulses emitted by a RRAT, the likelihood is a series of Bernoulli trials for each pulse emitted. This can be written as
\begin{equation}
  L(N,\phi_1 ... \phi_N) \propto \prod_{i=1}^{N} \phi_{i}^{d}(1-\phi_{i})^{1-d}
\end{equation}
\noindent where $L$ is the likelihood, $i$ is the pulse number, $N$ is the total (but unknown) number of pulses emitted by the RRAT, $\phi_i$ is the probability of detecting pulse $i$ and $d=1$ if a detection was made, or $d=0$ otherwise. Removing the terms in the product that go to one (e.g., when $d=1$, then the terms $(1-\phi_i)^{1-1}=1)$, we can rewrite this as
\begin{equation}
  L(N,\phi_1 ... \phi_N)\propto \bigg(\prod_{i=1}^{n}\phi_{i}\bigg) \bigg(\prod_{j=n+1}^{N}(1-\phi_{j})\bigg)
\end{equation}
where $n$ is the number of detected pulses. The first product results from all the pulses detected by the telescope, and the second product results from all those not detected by the telescope. In practice, $\phi_{i}$ depends on the pulse's brightness and the intrinsic luminosity function of the RRAT. Therefore,
\begin{equation}
  \phi_{i}(S_{i,det},\vec{\alpha})=P(det|S_{i,det})\int P(S_{i,det}|S_T,\sigma_d)P(S_T|\vec{\alpha}) \text{d}S_T
  \label{eq:detect}
\end{equation}
where $S_{i,det}$ is the detected brightness of pulse $i$. $S_T$ is the true intrinsic brightness as emitted by the RRAT. We emphasize the difference between $S_{T}$ and $S_{det}$ where $S_{det}$ is the detected brightness and comes with the errors due to the stochastic background noise of the telescope. $S_T$ is the intrinsic brightness of the pulse emitted from the source without any noise applied to it. The latter, even if inherently low, can be amplified by random noise fluctuations, rendering it detectable \red{and vice-versa}. Expressed differently, $S_{det}=S_{T}+S_{\sigma}$ where $S_{\sigma}$ is the stochastic noise of the data. \red{Depending on the chosen $S_{i,det}$ cutoff this effect can be non-negligible. Figure \ref{fig:injection_fn} shows that this effect is most apparent at the turn-off locations, near detection fractions of 0 and 1.} \injdet{} encodes all the selection biases of the telescope and detection pipeline. $P(S_{T}|\vec{\alpha})$ is the probability density of the intrinsic luminosity distribution of the RRAT where $\vec{\alpha}$ is a vector containing the parameters of the luminosity model, that is, $\vec{\alpha}=[\mu,\sigma]$, or $\vec{\alpha}=[k]$ for a log-normal or exponential distribution, respectively. This is discussed in further in section~\ref{sec:lum_dist}. \err{} designates the Gaussian detection error with $\sigma_d$ being the known and fixed error. $\sigma_d$ is measured empirically by performing injections, as discussed in Section \ref{sec:injections}. Due to $S_T$ being an unmeasurable parameter, we marginalize over it. A computational efficiency boost is achieved by realizing that \err{} is a Gaussian and $P(S_{det}|S_T,\sigma_d) = P(S_{det}-S_T|0,\sigma_d)$ allowing the right-hand integral of equation \ref{eq:detect} to be written as a convolution.

For pulses that aren't detected, we assume that the probability of non-detection is the same for every pulse. That is,
\begin{equation}
  1-\phi_{j}(\vec{\alpha}) = 1-\int P(det|S_{T})P(S_{T}|\vec{\alpha}) \text{d}S_{T}.
\end{equation}
This assumption must be made as there is no way to know the brightness of a pulse that is not detected. Finally, as the pulses are assumed to be independent, the order does not matter. The independence assumption does not consider refractive scintillation or the clustering of RRAT pulses. Refractive scintillation can change the overall shape of $P(S_T|\vec{\alpha})$ \red{and occurs on timescales of weeks \citep{10.1098/rsta.1992.0090}. Therefore, when applying LuNfit to data that spans longer than a few weeks, we point out that the fitted luminosity function will be the intrinsic luminosity function modulated by refractive scintillation.} However, carefully accounting for the effects of refractive scintillation \red{with different classes of luminosity functions} is out of the scope of this study. \red{We note that in this study, the timescales for observations in Section \ref{sec:results} span months and therefore, the luminosity function that has been fit is likely modulated by refractive scintillation. We do not make any attempts to remove these effects. Despite this, as seen in Section \ref{sec:results}, it is unlikely that the shape of the luminosity function is much different from a log-normal or exponential distribution.} Any clustering of RRAT pulses should not cause a problem for LuNfit as there is no dependence on each pulse's arrival time. We assume that all pulses from each cluster conform to a global luminosity distribution and do not consider the case where each cluster possesses an independent luminosity distribution. Thus, the likelihood is multiplied by ${N \choose n}$ and becomes
\begin{equation}
  L(\vec{\alpha},N|n,S_{det}) = {N \choose n}
  \bigg(1-\int P(det|S_{T})P(S_{T}|\vec{\alpha})\text{d}S_{T}\bigg)^{N-n}
    \prod_{i=1}^{n} P(det|S_{i,det})\int P(S_{i,det}|S_T,\sigma_d)P(S_T|\vec{\alpha}) \text{d}S_T.
  \label{eq:likelihood}
\end{equation}

For this analysis, we take a Bayesian approach utilizing nested sampling \citep{https://ui.adsabs.harvard.edu/abs/2004AIPC..735..395S} through the \texttt{dynesty} package \citep{https://ui.adsabs.harvard.edu/abs/2020MNRAS.493.3132S}\footnote{https://dynesty.readthedocs.io/en/stable/}. \red{Bayesian nested sampling is a method which breaks up the posterior into many nested ``slices'', assigns weights and volumes to each slice and recombines it again to form the posterior and evidence. It is a distinct way to approach Bayesian analysis compared to Markov Chain Monte Carlo and possesses multiple benefits. For example, nested sampling can sample the whole parameter space and avoid problems with multimodal distributions. By extension, obtaining the model evidence (marginal likelihood) for model comparisons through the Bayes ratio is trivial. Both these features are crucial for LuNfit.}

We use the detected signal to noise (S/N) for $S_{i,det}$. The strict definitions are given in section~\ref{sec:detections}. Nevertheless, other brightness metrics, like detected flux density or fluence, can be employed instead. Details of changing to flux density units are given in section \ref{sec:conversion}.
\subsection{Underlying Luminosity Distributions}\label{sec:lum_dist}
One caveat in this formulation is the luminosity distribution chosen to describe the detected single pulses, where choosing the incorrect model can significantly alter the luminosity and nulling characteristics of the source in question. However, as we employ a Bayesian nested sampling approach, we use the Bayes ratio to differentiate between different models. In section \ref{sec:simulations}, we demonstrate with simulations that the Bayes ratio is robust in general. For this reason, we have selected two models to describe pulsar emission characteristics. The first model is the log-normal distribution parameterized by the location, $\mu$, and scale, $\sigma$, described by

\begin{equation}
P(S_T|\mu,\sigma) = \frac{1}{S_T\sigma\sqrt{2\pi}}\exp\left(-\frac{(\ln S_T-\mu)^2}{2\sigma^2}\right).
\label{eq:logn}
\end{equation}

The second model is the exponential distribution parameterized by $k$, given by:

\begin{equation}
P(S_T|k) = k\exp(-kS_T).
\label{eq:exp}
\end{equation}

We chose these two models because they have demonstrated their ability to explain various neutron star single pulse emission phenomena, such as RRAT pulses \citep{10.1111/j.1365-2966.2011.18521.x,10.1093/mnras/sty1785,10.3847/1538-4357/aaee7b,10.1017/pasa.2019.30,10.3847/1538-4357/ac1da6,10.1093/mnras/stad2012}, giant pulses \citep{
10.1093/mnrasl/slz140,10.1051/0004-6361/200913729,10.1093/mnras/sty3058}, and regular single pulses \citep{10.1111/j.1365-2966.2012.20998.x,10.1093/mnras/stv397,10.1051/0004-6361/202142242}. Repeating FRBs have shown double-peaked Gaussian distributions which could also be considered in future studies \citep{10.1038/s41586-021-03878-5} .
\subsection{Priors}\label{sec:priors}
\begin{table}[]
\centering
\caption{Prior types and parameters}
\label{tab:priors}
\begin{tabular}{llll}
Parameter & Prior Type & $p_1$                 & $p_2$                     \\\hline\hline
$\mu$     & Gaussian   & 0                 & 4           \\
$\sigma$  & Inverse Gamma    & 1.938       & 1                           \\
k         & Inverse Gamma    & 1           & 1 \\
N         & Uniform    & n                 &  N$_{rot}$ = ($T_{obs}$/Period)             
\end{tabular}
\end{table}
We assign general priors for both the luminosity function parameters and the number of pulses. These priors are given in Table \ref{tab:priors}. Unlike Markov Chain Monte Carlo methods that do not necessarily require normalized priors, we need the Bayes Ratio for model comparisons and thus need robust evidence estimates. The priors are chosen according to expected behavior. However, they do not restrict unexpected behavior. To achieve this, all priors do not have strict limits unless those limits are unphysical, such as in the case of $N$. \red{All chosen values on the priors have been chosen based on simulations of pulsars for CHIME, where the underlying unit is S/N. Users of LuNfit should adjust these values if they are using flux, fluence, or other brightness metrics.}\\
For $\mu$, we choose a Gaussian prior described by the following equation
\begin{equation}
    P(\mu) = \frac{1}{p_2\sqrt{2\pi}}\exp{\bigg(-\frac{(\mu-p_1)^2}{2p_2^2}\bigg)}
\end{equation}
such that the prior contains $\sim68$\% of the probability mass between $-4$ and $4$. This ensures little bias towards particularly faint or bright pulsars. \red{Note that $\mu$ has units of ln(S/N), such that -4 corresponds with S/N$\approx$0.02 and 4 corresponds to S/N$\approx$54.60. Therefore, we allow a wide range of median values. If the user knows that the pulsar of interest is particularly bright/faint relative to the telescope sensitivity, the prior on $\mu$ should be adjusted to reflect that prior knowledge.}

For $\sigma$ and $k$ we choose an inverse gamma distribution as the prior. They are described by
\begin{equation}
    P(\alpha) = \frac{p_2^{p_1}}{\Gamma(p_1)}(1/\alpha)^{p_1+1}\exp({p_2/\alpha})
\end{equation}
These priors ensure support for $\sigma$ and k across the viable parameter range between 0 and $\infty$. However, we choose $p_1$ and $p_2$ such that for $\sigma$, $\sim$90\% of the probability mass is between 0 and 2, and for k, $\sim$90\% of the probability mass is between 0 and 10. The ranges represent the most likely parameter space for detectable pulsars. \red{Since $\sigma$ is unitless, we believe that the priors chosen for $\sigma$ should be near universal for all pulsars. We note that the prior on $\sigma$ did not need to be altered in any set of simulated parameters. Although k is also unitless, k determines the mean of an exponential distribution. Like $\mu$, the chosen prior for k can differ if the pulsar is particularly bright/faint relative to the telescope sensitivity.}

The prior for N is determined by the number of detections made and the amount of time we have observed the source. We assign a uniform distribution between $n$ and $N_{rot}=\lfloor T_{obs}/P \rfloor$ where $\lfloor \rfloor$ is the floor function, $T_{obs}$ is the amount of observation time, and $P$ is the pulsar period. If $P$ is unknown, we set $N_{rot}=\lfloor T_{obs}/1\text{ms} \rfloor$.

\subsection{Discrete N}\label{sec:discrete_N}
Discrete parameters are non-trivial to implement in Bayesian nested sampling. As N is necessarily discrete, we implement a pragmatic approach approximating its discreteness. \texttt{Dynesty} draws samples from the inverse cumulative distribution function (CDF). Therefore, to ensure discrete values of N are drawn, we use the \texttt{scipy.stats.randint.ppf}\footnote{https://docs.scipy.org/doc/scipy/reference/generated/scipy.stats.randint.html} function. This applies the following CDF:
\begin{equation}
    {\rm CDF} = \frac{\lfloor N \rfloor -n+1}{N_{rot} -n}
\end{equation}
While this implementation may cause errors if n and N are $\mathcal{O}(1)$, in the case of this study, both n and N and their associated errors are much greater than 1. Therefore, we do not expect significant effects on the inference of N.

\subsection{CHIME and detection of pulses}\label{sec:detections}
Here, we describe how we obtain $S_{i,det}$ for each of the detected pulses. To identify single pulses from known RRATs and pulsars, we first search the CHIME/Pulsar filterbank data using the CHIME/Pulsar Single-pulse PIPEline (CHIPSPIPE) \citep{10.1093/mnras/stad2012}. Once we detect a pulse, we extract 2 seconds of data around the Time of Arrival (TOA) and downsample it by a factor of 3, achieving a time resolution of approximately 1~ms. Next, we mask the channels contaminated by Radio Frequency Interference (RFI), dedisperse the data to the nominal DM of the pulsar, and average the data in frequency, producing a single time series per pulse. To remove the underlying baseline, we remove the pulse from the data and fit polynomials up to 10th order. For each fit, we calculate a reduced $\chi^{2}$ value and accept the fit where the reduced $\chi^2$ is closest to 1. This process is important as the CHIME/Pulsar baseline can be highly variable. Once the baseline is fit, we subtract it from the frequency collapsed time series, and the remaining background noise is described by the standard deviation of $X_{n}$, the baseline-subtracted, pulse removed time series ($\sigma_{noise}$).\\
To determine the amplitude of the pulse, we subtract the fitted baseline from the original 2-second chunk of data. Then, we perform a maximum likelihood fit of a Gaussian pulse given by the equation:
\begin{equation}
F(X)=A\exp\left(-\frac{1}{2}\frac{(X-X_{loc})^2}{W^2}\right)+B
\end{equation}
where $X_{loc}$ represents the location of the Gaussian peak, $X$ is the frequency collapsed time series, $A$ is the peak amplitude, $W$ is the Gaussian width, and $B$ accounts for any remaining DC offset in the time series. Throughout our study, unless otherwise specified, we use $S/N=S_{det}=\frac{A}{\sigma_{noise}}$. We perform the Gaussian pulse fit on each pulse detected by CHIPSPIPE. After fitting, we discard all $S_{det}<2$ to ensure that every pulse is a good fit to the Gaussian. This fiducial cut-off is also reflected in the selection effects that are measured.

\subsection{Injections}\label{sec:injections}

\begin{figure}
    \centering
    \includegraphics[width=0.5\linewidth]{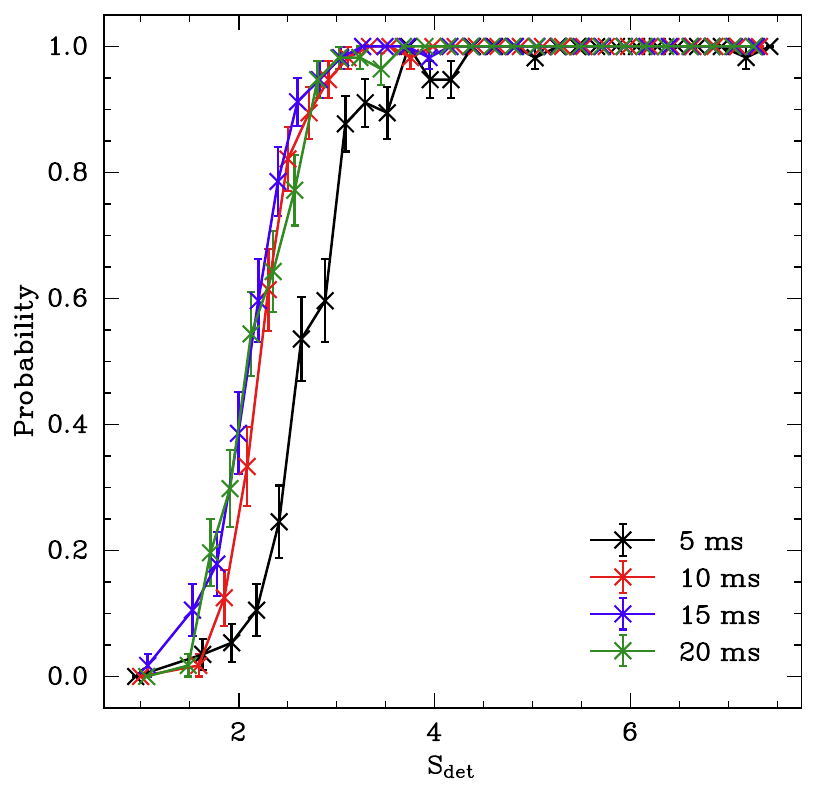}
    \caption{The width effects of CHIPSPIPE. The plot shows the probability of retrieval for different $S_{det}$ at varying widths. For regular slow pulsars, the Gaussian widths are generally greater than 10ms. Thus, they are not affected much by CHIPSPIPE selection effects. For sources that exhibit high levels of pulse-to-pulse width variability, one needs to model the width along with $S_{det}$ in LuNfit. For the current iteration of LuNfit, we only chose pulsars with Gaussian widths larger than 5~ms to avoid the sensitivity drop-off.}
    \label{fig:width_effects}
\end{figure}
\begin{figure}
    \centering
    \includegraphics[width=.49\linewidth]{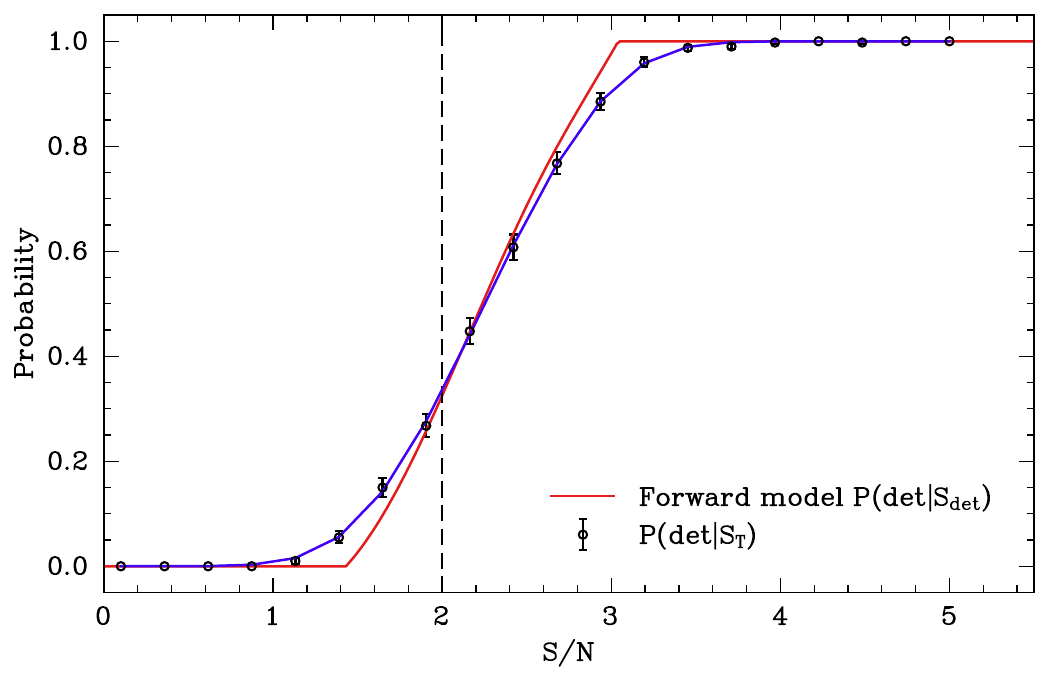}
    \includegraphics[width=.49\linewidth]{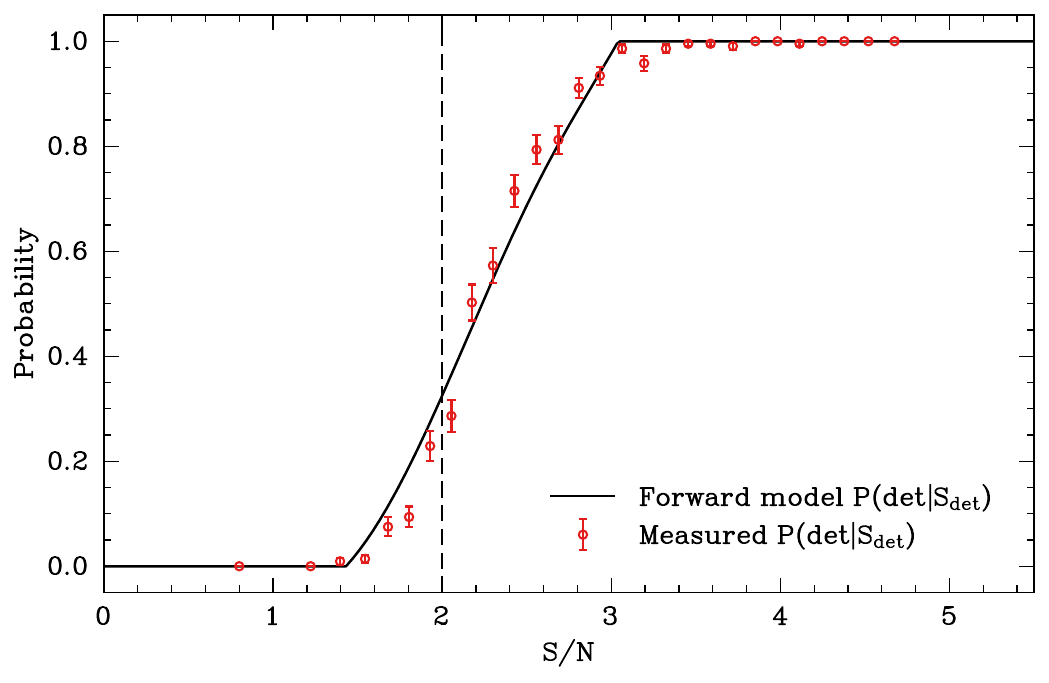}
    \caption{The selection effects of CHIPSPIPE. This plot shows the probability of detection given $S_{i,det}$ or $S_T$. The left shows the fitted \injt{} compared to the forward modeled \injdet{}. The right shows how the forward model \injdet{} compared to the measured values. This study uses the LuNfit implementation. We show the full selection effects here, but when performing LuNfit, we cut off the selection at $S_{i,det}=2$ to ensure robust pulses and retrieval probability measurements. The dotted line indicates this cut. The blue line is a best-fit polynomial for $P(det|S_T)$.}
    \label{fig:injection_fn}
\end{figure}
\begin{figure}
    \centering
    \includegraphics[width=0.5\linewidth]{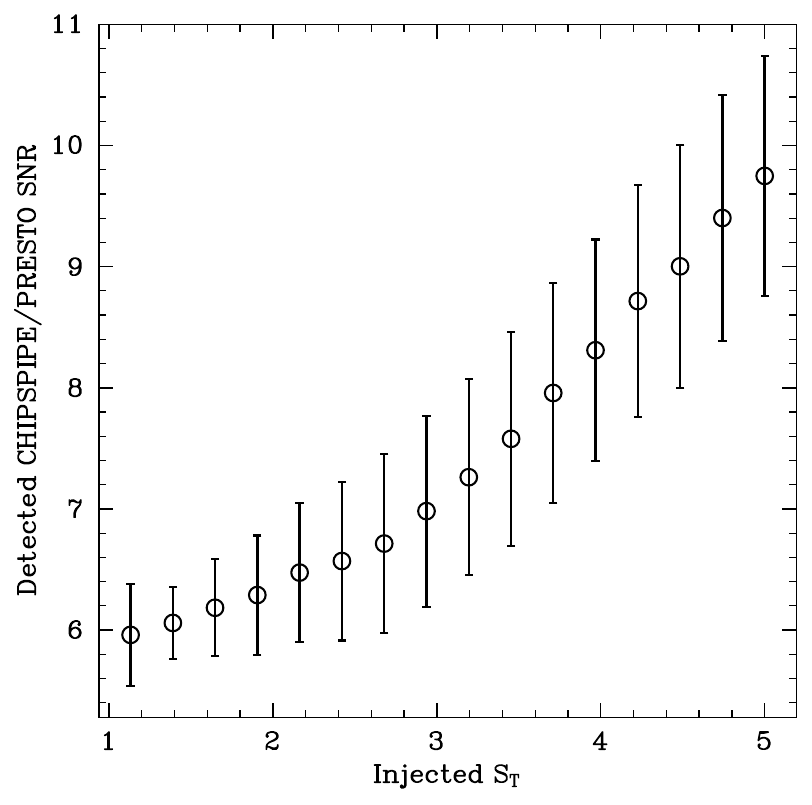}
    \caption{The detected CHIPSPIPE single pulse search S/N vs the injected $S_T$ value. Not all injected pulses are detected for detections of $S_T<\sim3$. Thus, the CHIPSPIPE S/N will be biased high as higher S/N pulses are being preferentially detected. This causes the flattening of the dependence at low $S_T$.}
    \label{fig:injvpresto}
\end{figure}

In this section, we describe the process of obtaining \injdet{}, \injt{}, and \err{}, with the primary goal of empirically determining the retrieval probability (or detection fraction) at different $S_{T}$ and $S_{i,det}$. These values are obtained empirically since the complexity of CHIME/Pulsar and CHIPSPIPE makes it impossible to derive them analytically. We inject simulated Gaussian pulses into real CHIME/Pulsar data. The simulated pulses are broadband with flat spectral indices and widths equal to the mean of the observed widths of the source in question. CHIPSPIPE is not sensitive to width changes for regular slow pulsars or RRATs, so injecting with one width per pulsar is sufficient. Figure \ref{fig:width_effects} shows the effects of width on pulse retrieval for CHIPSPIPE. We find that as long as most pulse widths remain above 10~ms then the width effect of CHIPSPIPE is minimal. Discussions of situations where one must include width effects in the likelihood are provided in more detail in section \ref{sec:limitations}. Since there are hundreds of hours of CHIME/Pulsar observations for each source spread across hundreds of observations, we spread out the injections over multiple observations to perform about 20,000 injections per pulsar. Spreading the injections across multiple observation epochs allows for an accurate description of day-to-day RFI variations.

For a given injection we find the amplitude of the injected pulse in the following way. We subtract the baseline by fitting a polynomial around each injection TOA, and we measure $\sigma_{noise}$ in the same way that is described in section \ref{sec:detections}. One can only control $S_{T}$ and not $S_{det}$ when injecting. Thus, we inject the desired $S_T$ based on the $\sigma_{noise}$. For each $S_{T}$, we inject multiple copies at different TOAs in different observation files. The injected ``observations'' are saved and processed using the same pipelines for real pulsar data. An example of the recovery rate is shown in Figure \ref{fig:injection_fn}, and binomial errors are assumed for each data point $\bigg(\sqrt{\frac{p(1-p)}{m}}\bigg)$, where $p$ is the measured recovery fraction and $m$ is the total number of injections at a specific $S_{T}$. The low $S_{T}$ values observed here are due to a matter of S/N definition, and Figure \ref{fig:injvpresto} illustrates the conversion between CHIPSPIPE/PRESTO detection significance and $S_{T}$.

We assume that \err{} is Gaussian, and as $S_T$ is a controlled parameter, $\sigma_d$ is the only unknown. To measure $\sigma_d$, we fit each injected pulse as described in section \ref{sec:detections}. Therefore for each group of injected $S_T$ we obtain a group of detected $S_{i,det}$. We then take the standard deviation of the retrieved $S_{i,det}$ for each of the three brightest values of $S_{T}$ to use for $\sigma_d$. Since \err{} is dominated by the stochastic noise in the data, we assume $\sigma_d$ is constant across all injected $S_T$. 

Two disparate methods are provided to measure \injdet{}. The first approach involves measuring $S_{i,det}$ directly from the injected ``observations''. For each injected pulse, we measure $S_{i,det}$ by the method outlined in section \ref{sec:detections}. We bin them and calculate the recovery fraction at each bin by comparing them with the detected injections. This approach is the most accurate but may not be feasible for systems like CHIME/FRB, where one cannot go back through the data stream to calculate $S_{det}$ for non-detections. However, we injected into filterbank data, and even non-detections can be fitted as we know the data segment where the injections were placed. This is shown by red circles in the right panel of Figure \ref{fig:injection_fn}.

In cases where the first approach is not feasible, a second approach is provided, where one can forward model based on \err{} and \injt{}. It is found that for CHIPSPIPE, \injdet{} can be well fit by a piecewise nth-order polynomial.

\begin{align}
        \begin{aligned}
      P(det|S_{det}) =
     \begin{cases} 
        0 \qquad \qquad  &S_d<S_{d0}\\
        c_0 + c_1 x_1 + c2 x_2^2 ... + c_n x_n^n \qquad\qquad  &S_{d0}<S_d<S_{d1}\\
        1 & S_d > S_{d1}
    \end{cases}
    \end{aligned} 
  \label{eq:piecewise_inj}
\end{align}
\noindent where $S_{d0}$ is determined by when $P(det|S_{det})$ first reaches 0 and $S_{d1}$ is when $P(det|S_{det})$ first reaches 1, enforcing the function to be continuous.\\
\begin{equation}
  P(det|S_{T})=\int P(det|S_{det}) P(S_{det}|S_{T})dS_{det}
  \label{eq:forward_model}
\end{equation}
Equation \ref{eq:forward_model} shows the relationship between the measured probabilities, $P(det|S_{T})$, and the derived property $P(det|S_{det})$. Forward modeling can be done by using the user's preferred fitting method. In this case,  we show a maximum likelihood fit for $P(det|S_{det})$ in Figure \ref{fig:injection_fn}. We perform many maximum likelihood fits with the polynomials of equation \ref{eq:piecewise_inj} ascending from first order to tenth order. The polynomial with the closest reduced $\chi^{2}_{r}$ to 1 is accepted. Therefore, the polynomial order can change for each set of injections depending on the best fit.

\section{Simulations}{\label{sec:simulations}}
To assess the limitations of LuNfit, we simulate various sets of pulsars with different luminosity parameters and numbers of detected pulses. For each set of luminosity parameters, we simulate a pulse, add Gaussian noise based on the measured \err{}, and then determine whether it will be detected according to the measured \injdet{}. This process is repeated until the desired number of pulses is obtained. We then employ LuNfit to characterize each set of simulations. In total, 20 different sets of parameters are simulated for each luminosity distribution.
We simulate one realization for each set of $\vec{\alpha}$ parameters for the following scenarios:
\begin{enumerate}
        \item    150 detections, which represents the minimum required for a reliable fit for LuNfit.
        \item    500 detections, a reasonable number typically observed for the most prolific RRATs discovered by CHIME \citep{10.1093/mnras/stad2012}.
        \item    1000 detections, approximately the number of detections made by the Five-hundred-meter Aperture Spherical Telescope (FAST) for the most prolific FRBs \citep{10.1088/1674-4527/ac98f8,10.1038/s41586-021-04178-8,10.48550/arXiv.2304.14665}.
        \item    5000 detections, The best case scenario that can be achieved for the most prolific RRATs.
\end{enumerate}
These simulation results are presented in \red{Appendix} Figure \ref{fig:lunfit_exp} and \ref{fig:lunfit_logn}. \red{Note that all simulations of the log-normal luminosity model were made with $\sigma=0.5$.} In summary, we recover the correct parameters in almost all situations, with the exponential model requiring fewer detections for the same error region compared to the log-normal model. \red{This is due to the fact that the log-normal distribution is characterized by two parameters (as opposed to one for an exponential distribution) which allows more flexibility when modelling the data. The recovered parameters are generally more accurate when there are more detected pulses and (or) the median luminosity is higher. The true N will be lower for a higher median luminosity, and LuNfit will not need to account for as many missed pulses, leading to smaller error regions. Conversely, for lower median luminosities, LuNfit accounts for more missed bursts, and the error regions are larger. This characteristic is universal across luminosity models.} These simulations allow us to understand the limitations of LuNfit and provide valuable insights into its performance under various scenarios.

We compare the exponential and log-normal luminosity distributions via the Bayes ratio to see if LuNFit can correctly identify the correct true underlying distribution. We simulate four scenarios for both the log-normal and exponential distributions: 150, 500, 1,000, and 5,000 detections \red{(that is, the true underlying distribution is log-normal or exponential)}. We simulate 20 parameter sets for each scenario. We vary $\mu$ from -0.5 to 1 for the log-normal model while keeping $\sigma=0.5$. For the exponential distribution, we vary k from 0.5 to 2. \red{Appendix} Figures \ref{fig:bayes_exp} and \ref{fig:bayes_logn} show the Bayes ratios between the log-normal and exponential models for all of our simulations. LuNfit can retrieve the correct luminosity distribution for most cases. In general, when a pulsar follows an exponential distribution, LuNfit can always choose the correct model. On the other hand, with only 150 detections, the Bayes ratio is always low for the log-normal distribution. The Bayes ratio for log-normal distributions increases as more detections are made or $\mu$ increases. The reason for the $\mu$ dependence is that log-normal distributions approach exponential distributions for low values of $\mu$. Therefore, it becomes more difficult to parse the two distributions apart.
\subsection{Validation on known pulsars}\label{sec:validation}
\begin{figure}
    \centering
    \gridline{\fig{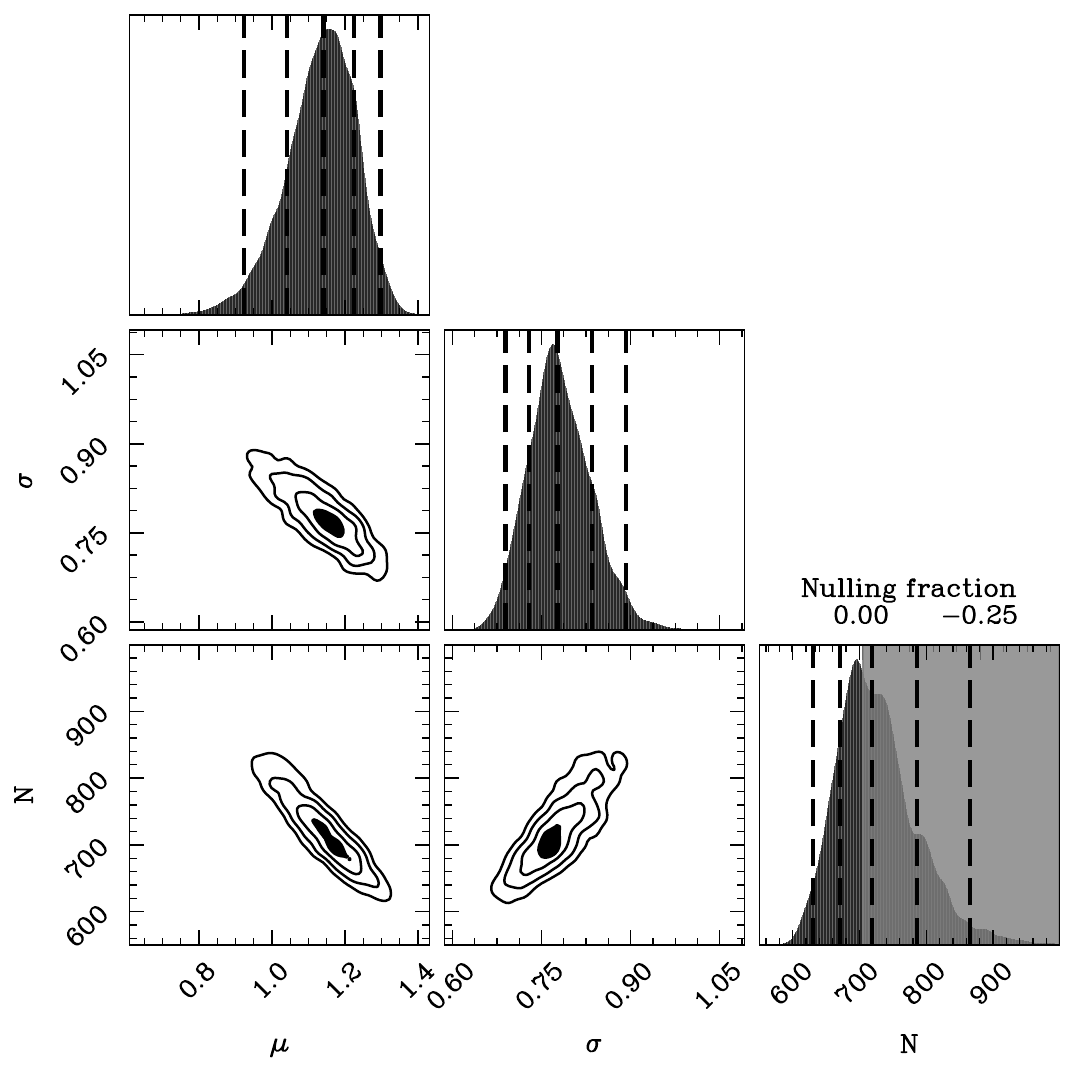}{.4\textwidth}{(a) B1905+39 without nulling fraction cut off}
    \fig{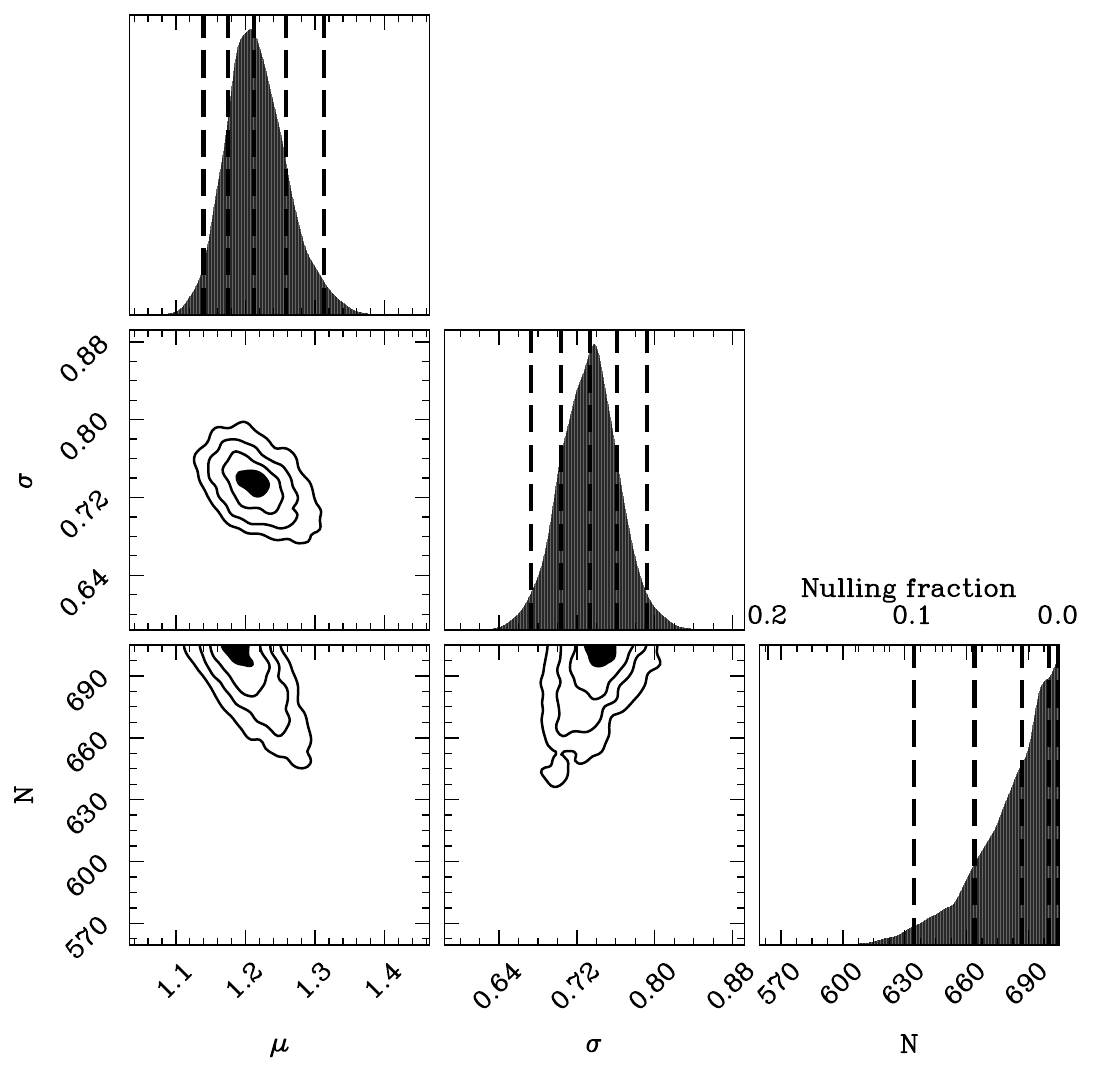}{.4\textwidth}{(b) B1905+39 with nulling fraction cut off at 0} }
    \gridline{\fig{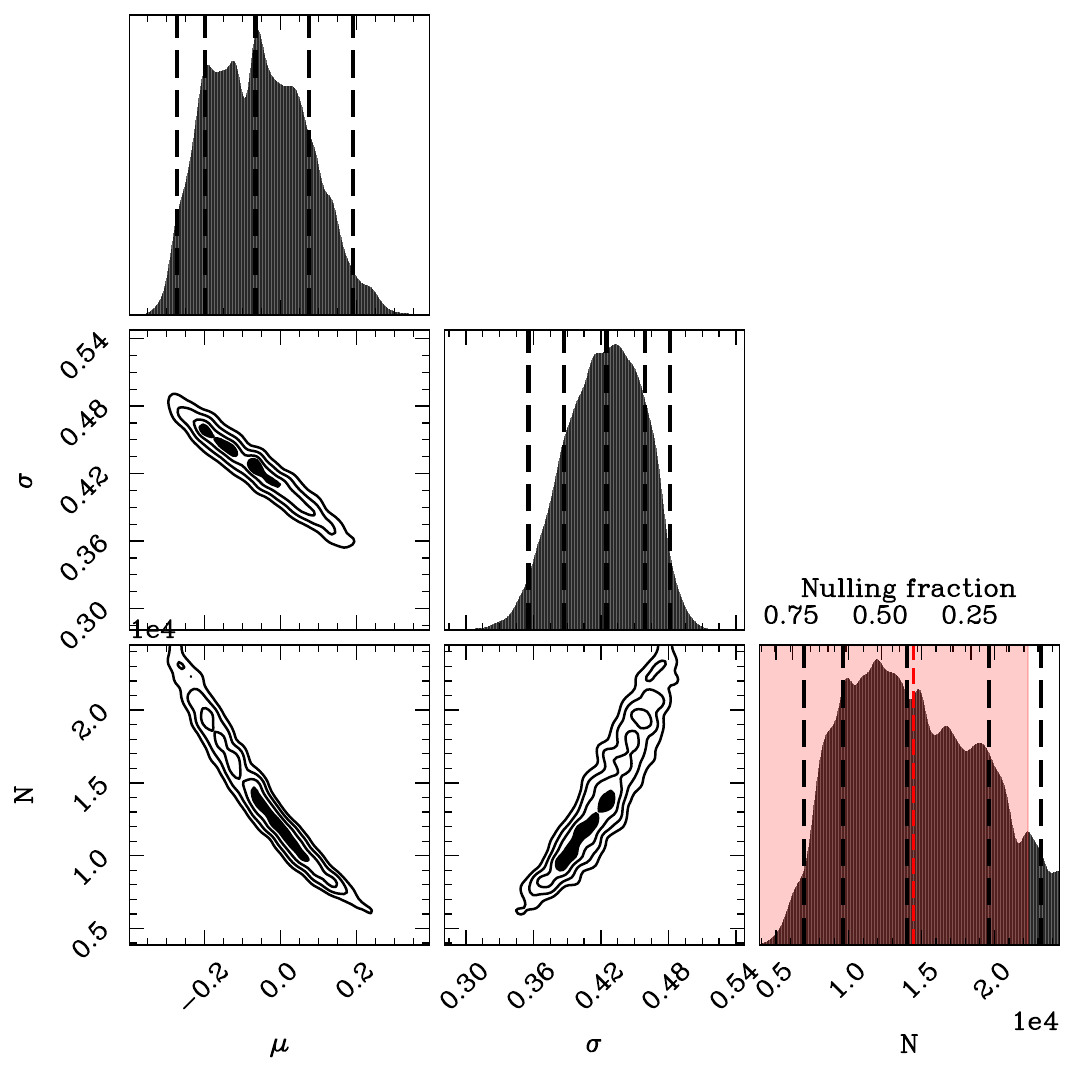}{.4\textwidth}{(c) J2044+4614} }
    \caption{(a) LuNfit for B1905+39, as expected, the nulling fraction peaks at $\sim0$. The shaded region is forbidden due to negative nulling fractions. We show this region to demonstrate that even when physical constraints are not set, LuNfit converges on reasonable values. (b) LuNfit for B1905+39 again, with the physical limits for nulling fraction. (c) LuNfit for J2044+4614. The predicted nulling fraction according to the method outlined in \cite{10.1093/mnras/176.2.249} is $\sim$0.41. This is shown by the red dashed line. The red-shaded region shows the allowed nulling fractions provided by \cite{10.3847/1538-4357/abb94f}. The black dashed lines show the 68th percentile, 95th percentile confidence intervals and the median.}
    \label{fig:validation}
\end{figure}
We validate our method by performing LuNfit on two pulsars with known nulling fractions, where the nulling fraction is given by $NF = 1-\frac{N}{N_{rot}}$ and $N_{rot}=\frac{T_{obs}}{P}$ is the total number of rotations that the observation time allows. This is shown in Figure \ref{fig:validation}. B1905$+$39 (P=1.236~s) is a bright, slow isolated pulsar and not known to exhibit any nulling. Indeed, we find no evidence of nulling in the LuNfit results given in Figure \ref{fig:validation}. We used one CHIME/Pulsar observation on MJD 59161, totaling 871~s, and detected 473 pulses. When LuNfit is applied to this pulsar, it accurately indicates a nulling fraction of $-0.02(5)$, consistent with 0 and confirming its non-nulling nature. In this calculation, we have lifted the prior on N and thus allow $NF<0$ to show the efficacy of LuNfit. This is shown in Figure \ref{fig:validation} (a) and (b). In Figure \ref{fig:validation} (a), we remove the constraints on the prior for N to show that LuNfit converges on the correct value even when not constrained. The constraints on N are then reapplied in Figure \ref{fig:validation} (b).

J2044$+$4614 (P=1.393~s), on the other hand, is a known nulling pulsar, and its nulling properties were previously discovered in \cite{10.3847/1538-4357/abb94f}. We used 39 CHIME/Pulsar observations from MJD 58808 to MJD 58579 and detected 585 bursts with 34130~s of observations. We show the LuNfit results in Figure \ref{fig:validation}. To compare our results with previously documented methods, we use the nulling measurement method discussed in \cite{10.1093/mnras/176.2.249}. The LuNfit derived nulling fraction is 0.4(2) and agrees with the \cite{10.1093/mnras/176.2.249} method value of $\sim0.41$. The two methods are shown in Figure \ref{fig:validation} (c). 
In addition, the nulling fraction as measured by \cite{10.3847/1538-4357/abb94f} is $>$9\%, consistent with our findings and shown as the shaded region in Figure \ref{fig:validation} (c).


\subsection{Conversion to flux density units}\label{sec:conversion}
Conversion from the S/N based $S_{det}$ to Jy-based units can be done by applying the single pulse radiometer equation \citep[e.g][]{10.3847/1538-4357/ac1da6},
\begin{equation}
    S_{peak} = S/N \frac{T_{sys}}{G\sqrt{\Delta \nu \Delta t~n_p}}
    \label{eq:single_pulse_radiometer}
\end{equation}
to each detection. Then one would repeat the analysis detailed in section \ref{sec:methods} with $S_{peak}$ in place of $S_{det}$, being careful to apply the same radiometer conversion factor to both the \injdet{} and \err{}. We show the results of this in Table \ref{tab:results}. Only the luminosity function is altered in this change of variables, and thus, the result for N does not change in Table \ref{tab:results}. We caution, however, that due to CHIME/Pulsar's poor flux density calibration issues \citep{10.3847/1538-4357/ac1da6,10.1093/mnras/stad2012}, these values should only be taken as a lower limit. 

\red{
An alternative method for converting the flux is to transform the resultant distribution.
We can directly transform the units of the luminosity distribution to flux density by
\begin{equation}
  P(S/N) \text{d}S/N = P(S_{peak}) \text{d}S_{peak}.
\end{equation}
Where $P(S/N)$ is the probability distribution of the S/N, and $P(S_{peak})$ is the probability distribution of the peak flux density. Rearranging and applying equation \ref{eq:single_pulse_radiometer} gives
\begin{equation}
    P(S_{peak}) = P(S/N) \bigg(\frac{T_{sys}}{G\sqrt{\Delta \nu \Delta t~n_p}}\bigg)^{-1}.
    \label{eq:flux_conversion_2}
\end{equation}
}
\section{Results}\label{sec:results}
\begin{figure}[ht]
    \centering
    \gridline{\fig{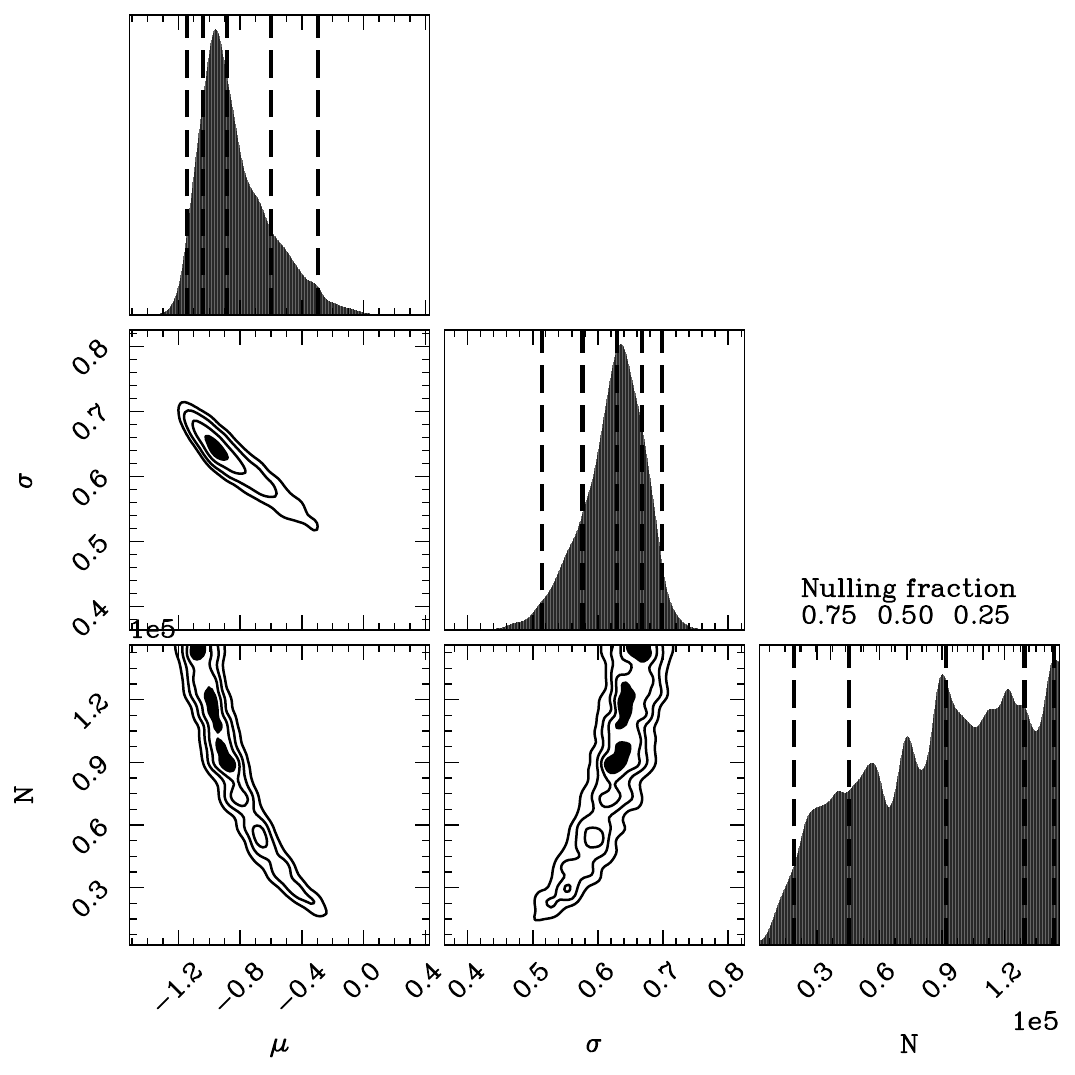}{0.4\textwidth}{(a)}
                \fig{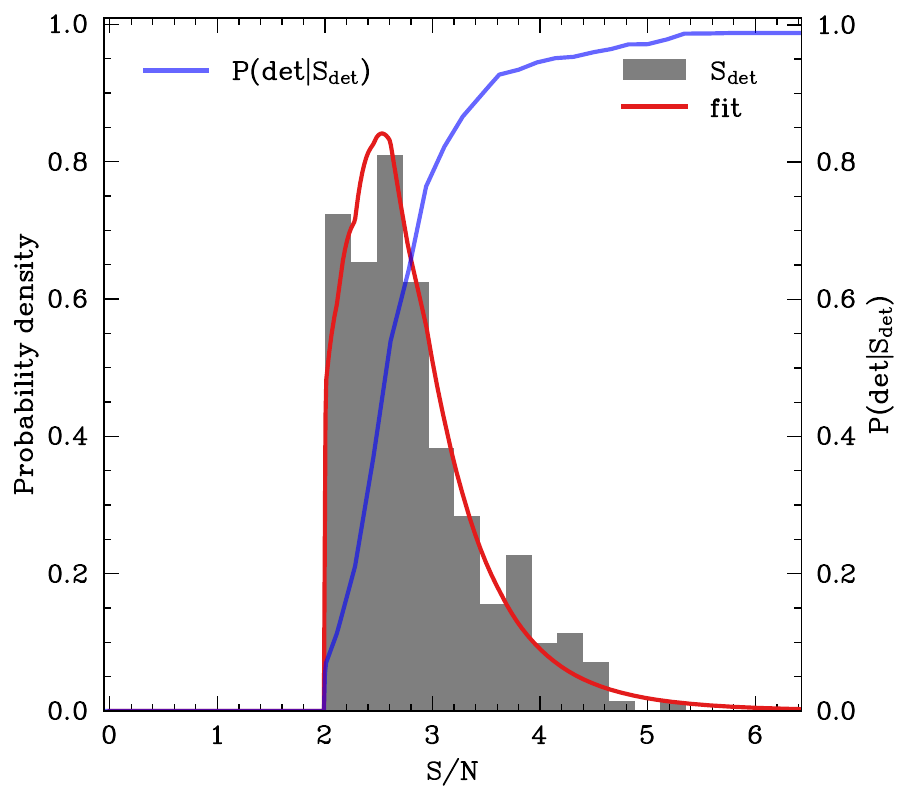}{0.4\textwidth}{(b)}
              }
    \gridline{\fig{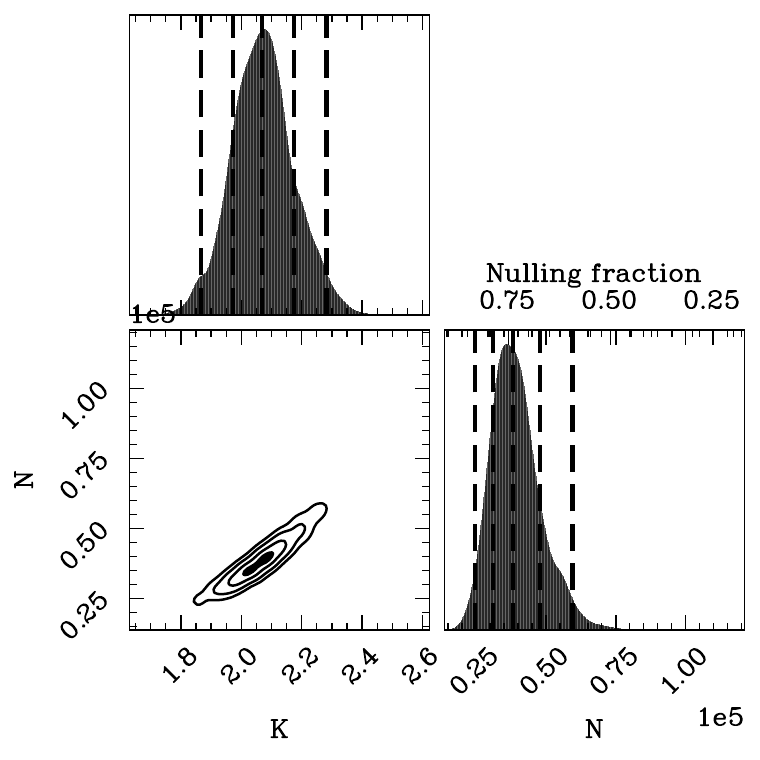}{0.4\textwidth}{(c)}
                \fig{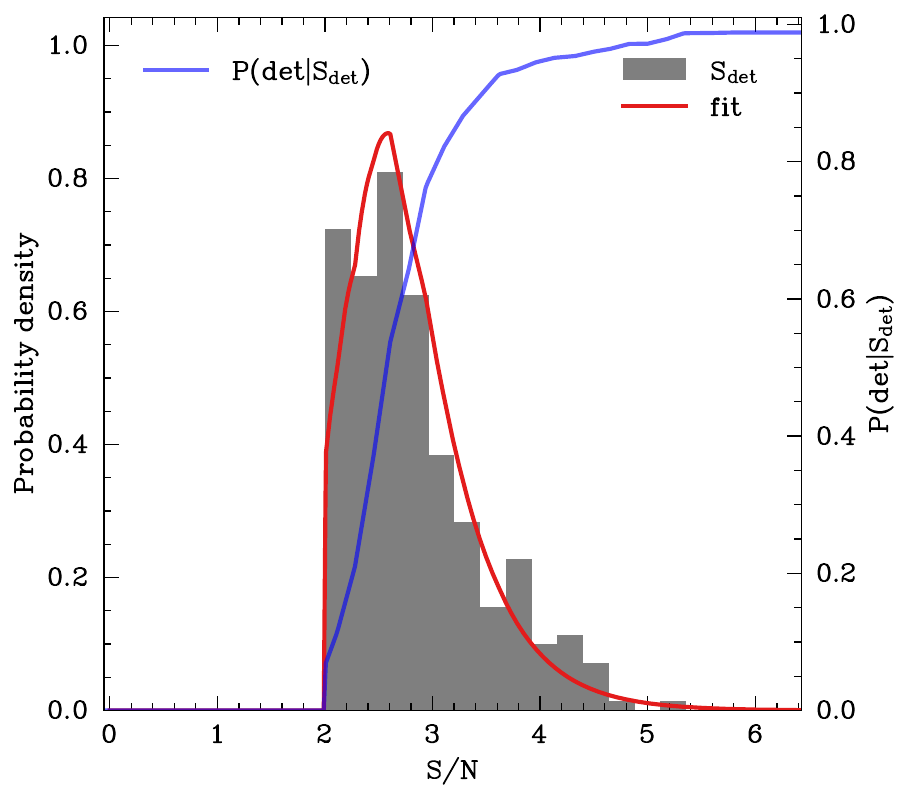}{0.4\textwidth}{(d)}
              }
    \caption{Corner plot and fit for J0012+5431. The log of the Bayes ratio is 0.7(3). (a) and (b) show the log-normal model, and (c) and (d) show the exponential model. Figures (a) and (b) show the LuNfit corner plots for J0012+5431. Figures (b) and (d) show the LuNfit fits as applied to the detections of J0012+5431. The black dashed lines show the 68 percentile, 95 percentile, and the median. We also provide the selection effect for this source as the blue line.}
    \label{fig:results_J0012}
\end{figure}

\begin{figure}[ht]
    \centering
    \includegraphics[width=.4\linewidth]{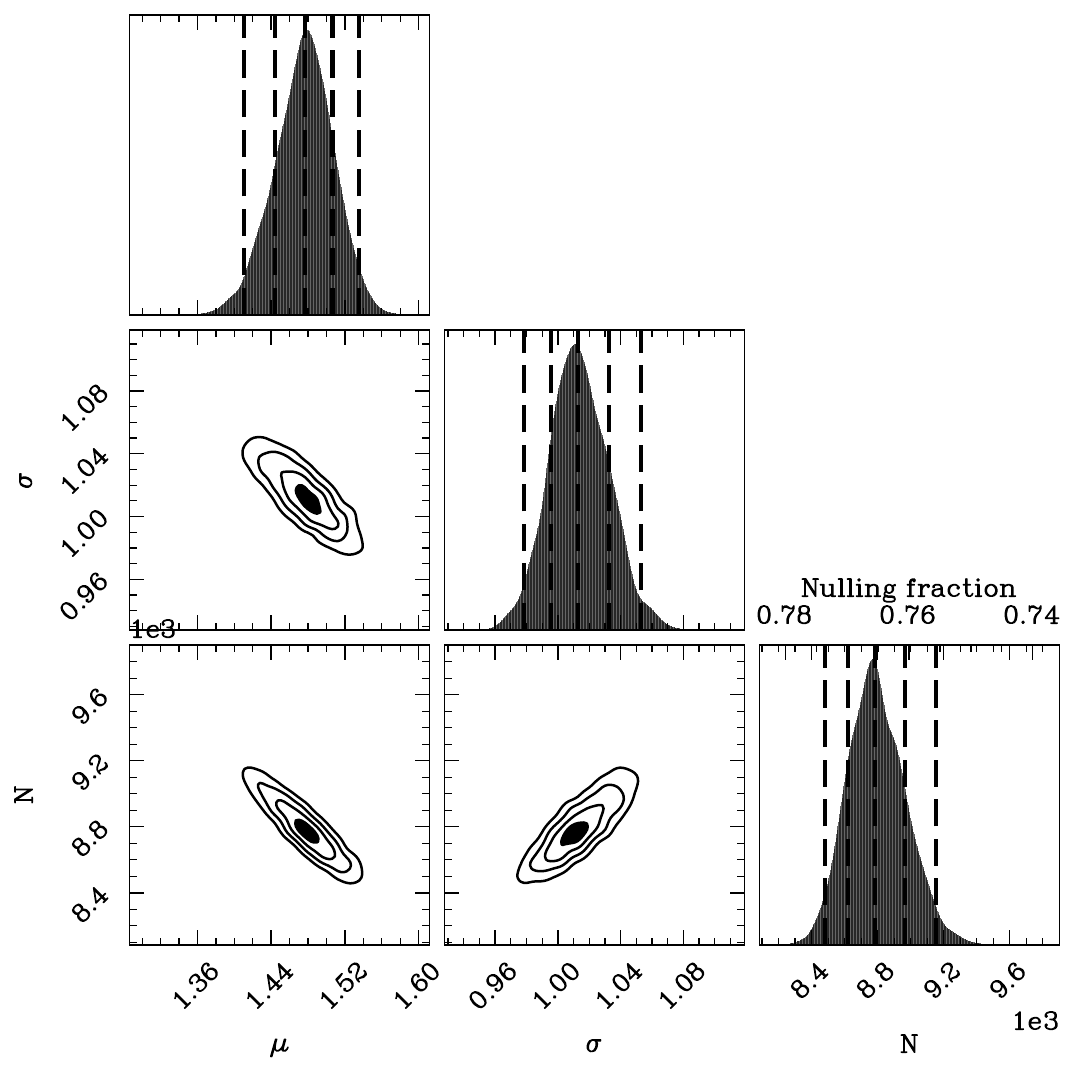}
    \includegraphics[width=.4\linewidth]{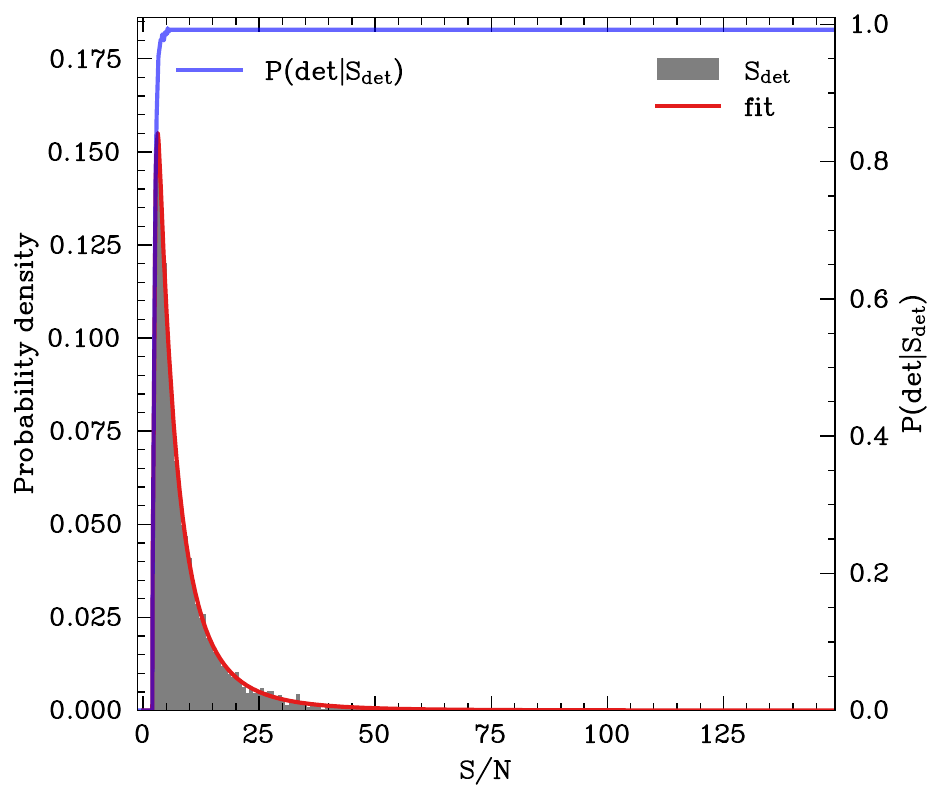}
    \caption{Corner plot and fit for J1538+2345. As J1538+2345 highly prefers the log-normal model, only that is shown. The left panel shows the log-normal constraints on J1538+2345, and the right panel shows the fit given the mean values. The black dashed lines show the 68 percentile, 95 percentile, and the median. }
    \label{fig:results_J1538}
\end{figure}

\begin{figure}[ht]
    \centering
    \gridline{\fig{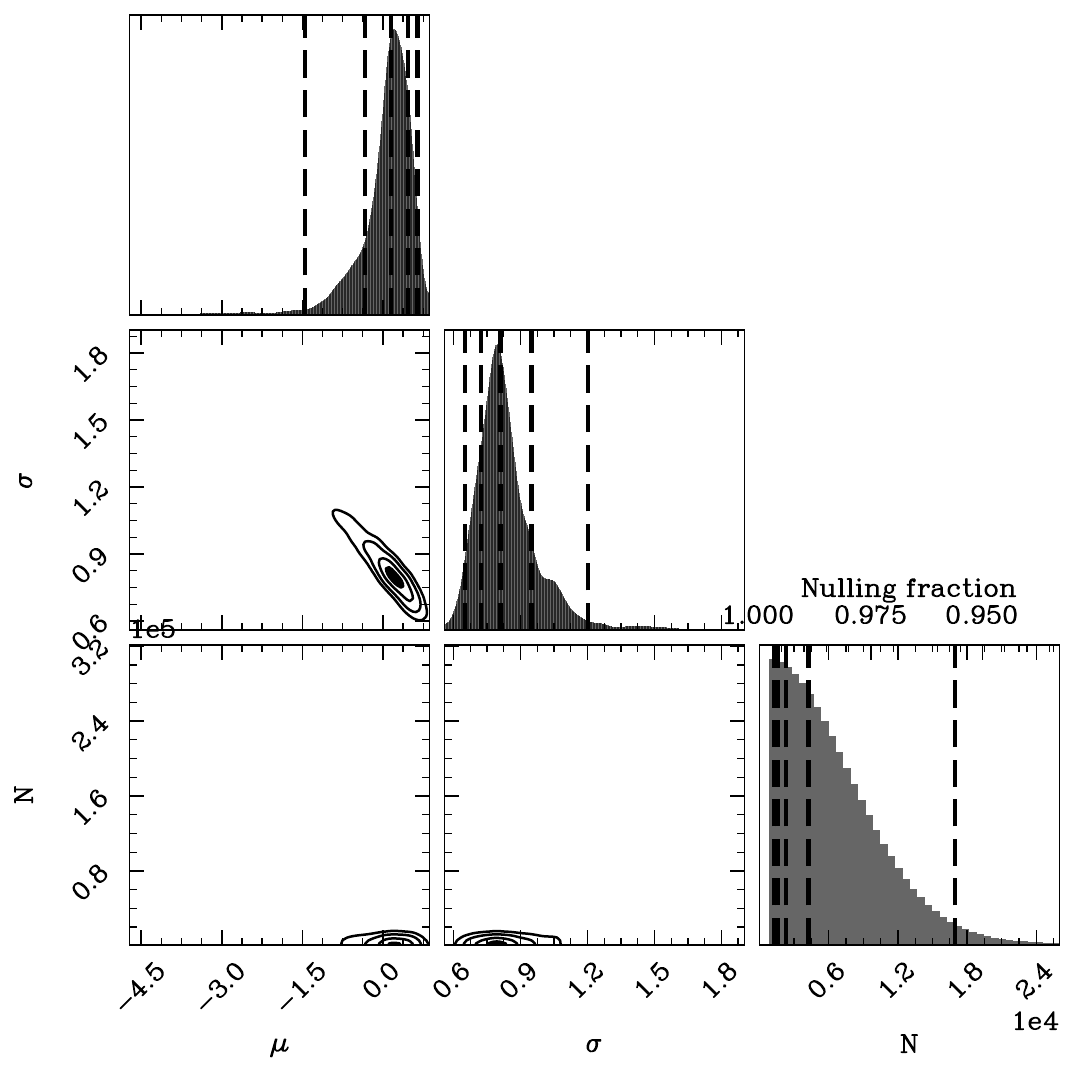}{0.4\textwidth}{(a)}
                \fig{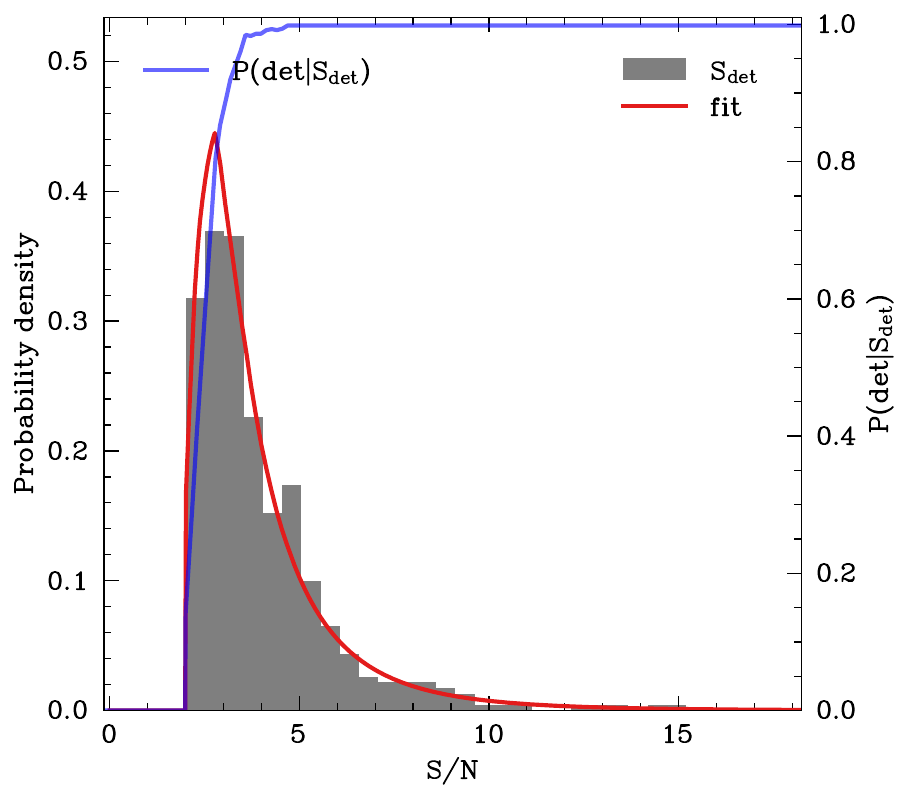}{0.4\textwidth}{(b)}
              }
    \gridline{\fig{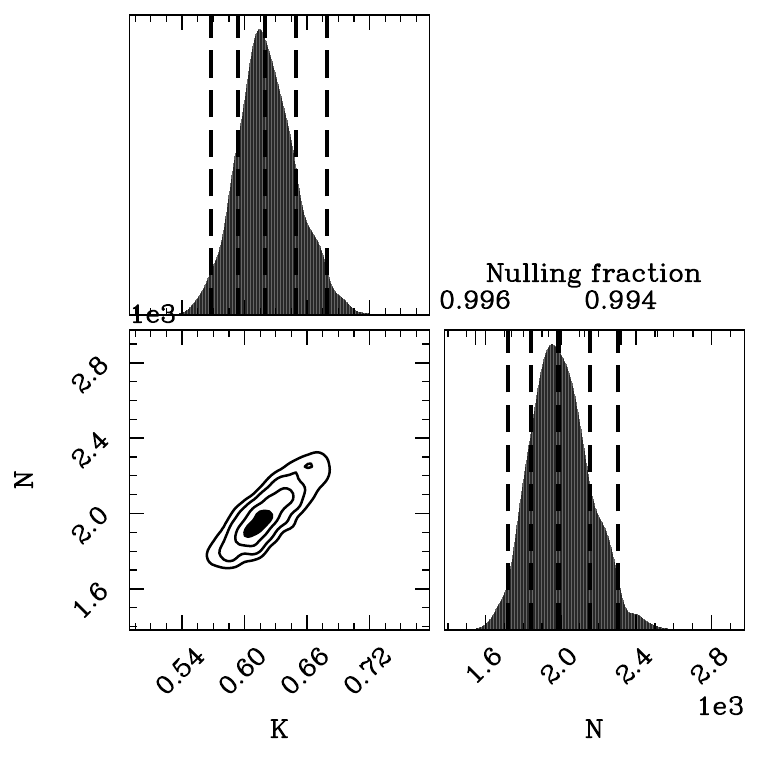}{0.4\textwidth}{(c)}
                \fig{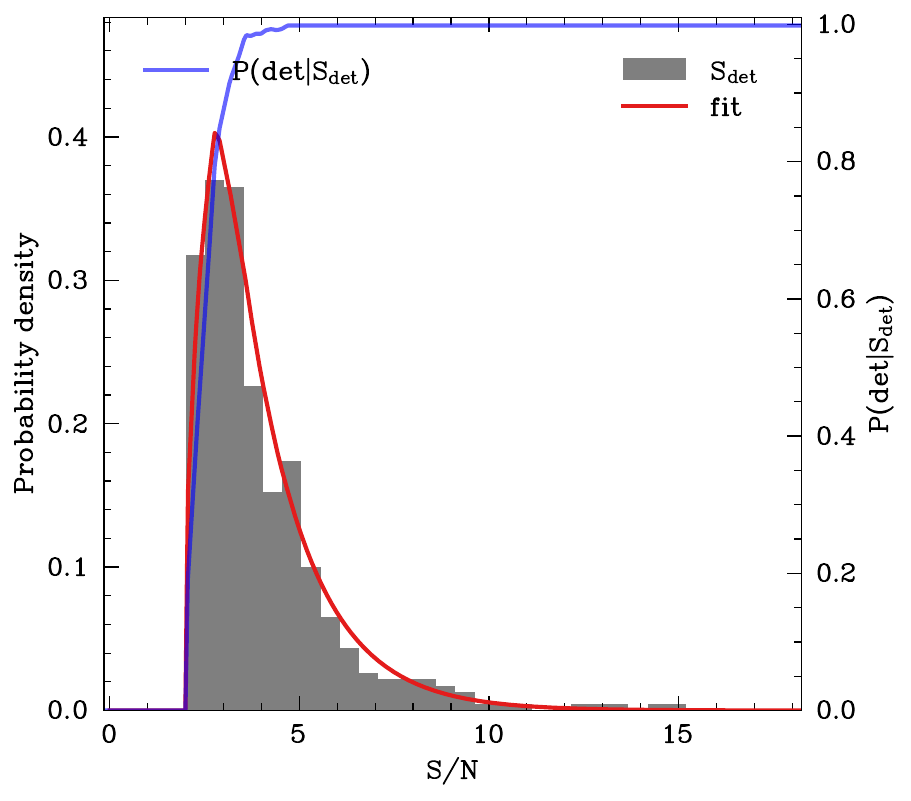}{0.4\textwidth}{(d)}
              }
    \caption{Fit and corner plots for J2355+1523. The log of the Bayes ratio is 2.7(4). (a) and (b) show the log-normal model, while (c) and (d) show the exponential model. Figures (a) and (b) show the LuNfit corner plots for J2355+1523. Figures (b) and (d) show the LuNfit fits as applied to the detections of J2355+1523. The black dashed lines show the 68 percentile, 95 percentile, and the median. We also provide the selection effect for this source as the blue line.}
    \label{fig:results_J2355}
\end{figure}

\begin{table}[]
\centering
\caption{Table of results for J0012$+$5431, J1538$+$2345 and J2355$+$1523. While both exponential and log-normal fit parameters are provided, \red{Two of the three} RRATs favor the log-normal model. Whereas J0012$+$5431 only weakly favors the log-normal distribution, J1538$+$2345 is has a strong preference for the log-normal distribution. \red{The only RRAT to favor the exponential model is J2355+1523, but it is a weak Bayes ratio with large errors. In either case, both distributions for J2355+1523 strongly suggest a high nulling fraction.} All the uncertainties provided are the 68th percentile confidence intervals.}
\label{tab:results}
\begin{tabular}{llll}
Pulsar                      & J0012+5431   & J1538+2345 & J2355+1523 \\\hline\hline
\# of detections            & 194          & 6272       & 453        \\
$\mu$                       & -0.8(2)      & 1.48(3)    & 0.1(3)    \\
$\sigma$                    & 0.62(5)      & 1.01(2)    & 0.8(1)    \\
$N_{logn}$                  & 88000(36000) & 9400(200)  & 2300(1900) \\
k                           & 2.1(1)       & 0.141(2)   & 0.62(3)    \\
$N_{exp}$                   & 39032(8700)  & 8910(70)   & 2010(160)  \\ \hline\hline
$\mu_{flux}$                & -4.0(2)       & -1.94(3)    & -3.1(6)     \\
$\sigma_{flux}$             & 0.62(5)      & 1.01(2)    & 0.8(1)     \\
$k_{flux}$                  & 51(5)     & 4.31(6) & 16(1)   \\ \hline\hline
Observation Time (s)        & 442410       & 129196     & 352933     \\
Period (s)                  & 3.025        & 3.449      & 0.913      \\
Nulling Fraction            & 0.4(2)       & 0.749(5)   & 0.995(3)   \\
$\ln{Z_{ln}}-\ln{Z_{exp}}$ & 0.7(3)       & 299.3(4)   & -0.6(5)     \\

\end{tabular}
\end{table}
 We present our findings concerning the constrained luminosity function, the intrinsic nulling fractions, and the total pulse number for three RRATs discovered in
  \cite{10.1088/0004-637X/809/1/67} and \cite{10.1093/mnras/stad2012}. They are J0012+5431, J1538+2345 and J2355+1523. Detailed results are summarized in Table \ref{tab:results}, while visual representations of the \texttt{dynesty} nested sampling fits can be found in Figures \ref{fig:results_J0012}, \ref{fig:results_J1538}, and \ref{fig:results_J2355}.

Upon examination, a notable trend emerges; the fitted exponential distribution consistently exhibits a higher median $S_T$ than the log-normal distribution, encompassing all pulsars outlined in Table \ref{tab:results}. This distinction consequently results in a diminished total pulse count, N, for the exponential distribution.\\
J0012+5431 is a RRAT first discovered in \cite{10.1093/mnras/stad2012}. As a result, we had an abundant amount of CHIME/Pulsar data available for this analysis. The LuNfit results for J0012+5431 are presented in Figure \ref{fig:results_J0012}. While it initially appears to be extremely sporadic ($\sim 1.6$ pulses per hour), LuNfit shows that this source is actually very faint and potentially less sporadic than initially thought. Consequently, CHIME/Pulsar appears to be detecting only the most intensely luminous tail of the distribution. We folded multiple observations to find signs of continuous emission without success. However, this is in line with expectations, considering the source's faintness, prolonged rotational period, and the brevity of CHIME/Pulsar observation windows. The long rotational period dictates that within the brief $\sim10$-minutes of a CHIME/Pulsar observation, accumulating a substantial number of rotations to bolster the signal-to-noise ratio significantly is unattainable. This \red{suggests} that follow-up with a more sensitive instrument with longer integration times such as FAST or the GBT could confirm our findings. With the Bayes Ratio only slightly leaning in favor of the log-normal model, we present LuNfit's results for both the log-normal and exponential models in Figure \ref{fig:results_J0012}.\\

J1538+2345 is a RRAT first discovered in \cite{10.1088/0004-637X/809/1/67}. For this RRAT, LuNfit heavily favors the log-normal distribution. Thus, only that is provided in the LuNfit results in Figure \ref{fig:results_J1538}. This source is extremely prolific for a RRAT with a nulling fraction of 0.749(5). As many telescopes have observed this source, we compare our results to that of the observational burst rates reported by the Low Frequency Array (LOFAR) \citep{10.1088/0004-637X/809/1/67}, the Green Bank Telescope (GBT) \citep{10.1088/0004-637X/809/1/67} and the Five Hundred Meter Aperture Synthesis Telescope (FAST) \citep{10.1007/s11433-018-9372-7} who reported observational nulling fractions of 0.94, 0.93, and 0.71 respectively. The FAST observations align the best with our selection corrected nulling fraction of 0.749(5). This is likely due to the extreme sensitivity of FAST, enabling it to probe the intrinsic properties of J1538+2345. We obtained a much higher number of observations than \cite{10.1007/s11433-018-9372-7}. Therefore, we likely have a more accurate long-term average. The slight discrepancy between the LuNfit results and the FAST results is likely due to the length of observations. The study by \cite{10.1007/s11433-018-9372-7} was conducted with only four observations, each lasting 30 minutes. The initial two observations yielded significantly elevated levels of emitted pulses from J1538+2345, which likely contributed to the lower nulling fractions reported by FAST.

J2355+1523 is a RRAT first discovered in \cite{10.1093/mnras/stad2012}. Among the RRATs in that study, it was one of the most prolific sources identified, making it an ideal candidate for LuNfit. The LuNfit results are shown in Figure \ref{fig:results_J2355}. We find that the Bayes ratio marginally favors the exponential distribution. Thus, we provide LuNfit results for both models in Figure \ref{fig:results_J2355}. In either case, the source is bright, and the predicted nulling fraction is close to 1. \\

\section{discussion}\label{sec:discussion}
We have shown through this study that LuNfit is an effective tool for probing the intrinsic luminosity function and burst rate of RRATs. This has been validated through simulation and comparison with pulsars of known nulling fractions. We then applied LuNfit to three known RRATs, finding their intrinsic luminosity functions and nulling fractions. Below, we provide a detailed discussion of LuNfit's limitations, how it compares with other techniques, and its use cases. Furthermore, we provide a discussion on how LuNfit can be adapted to be used for exotic radio transients such as FRBs and radio magnetars.

\subsection{Comparisons with other methods}\label{sec:comparison}
\begin{figure}
    \centering
    
    \includegraphics[width = .4\textwidth]{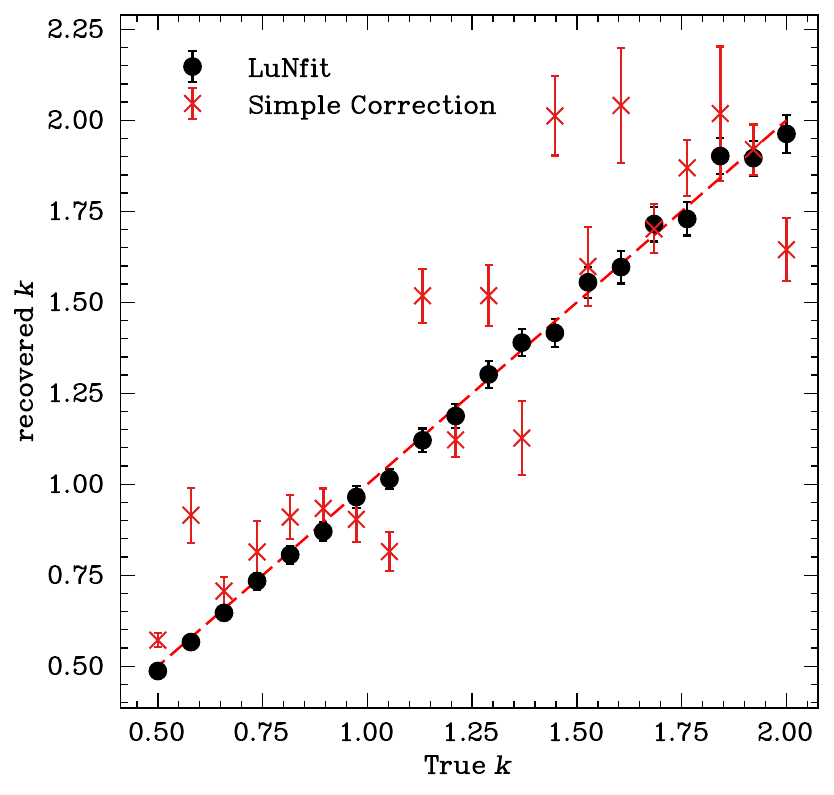}
    \includegraphics[width = .38\textwidth]{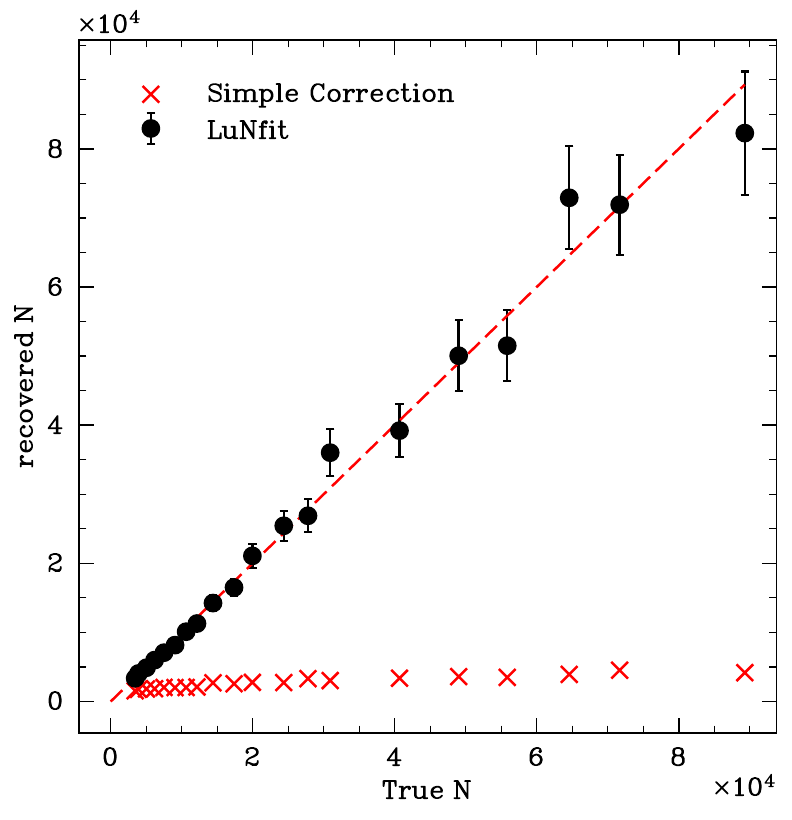}
    
    \caption{Comparison between a simple correction using the selection function and LuNfit for an exponentially generated distribution with 1000 detections. The "simple correction" is the method described in \cite{10.1038/s41586-021-03878-5}, and a full description is provided in section \ref{sec:comparison}.}
    \label{fig:compare-exp}
\end{figure}
\begin{figure}
    \centering
    
    \includegraphics[width = .3\textwidth]{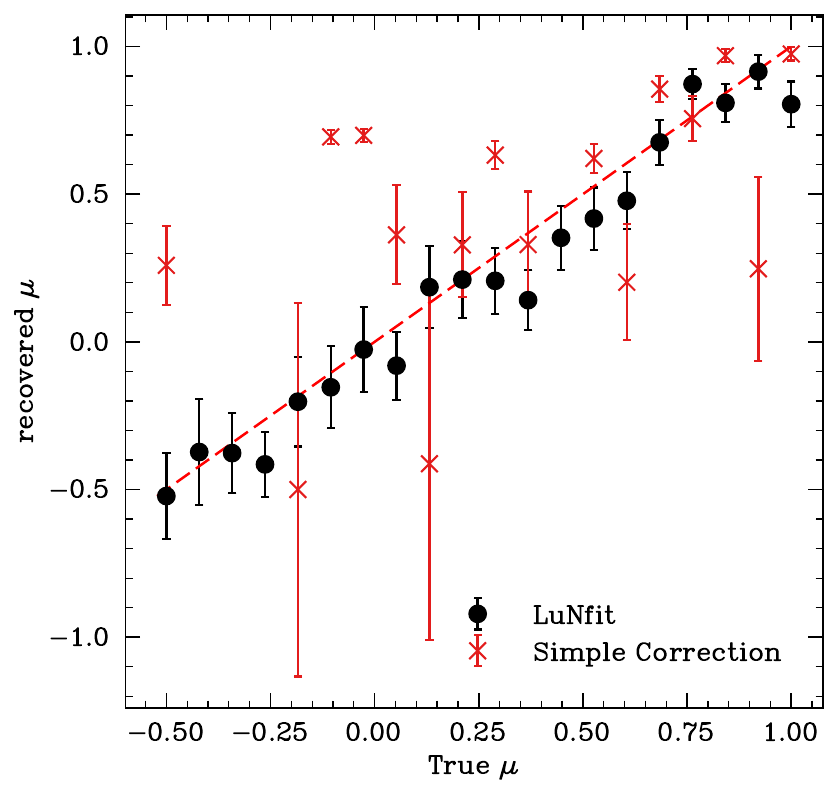}
    \includegraphics[width = .3\textwidth]{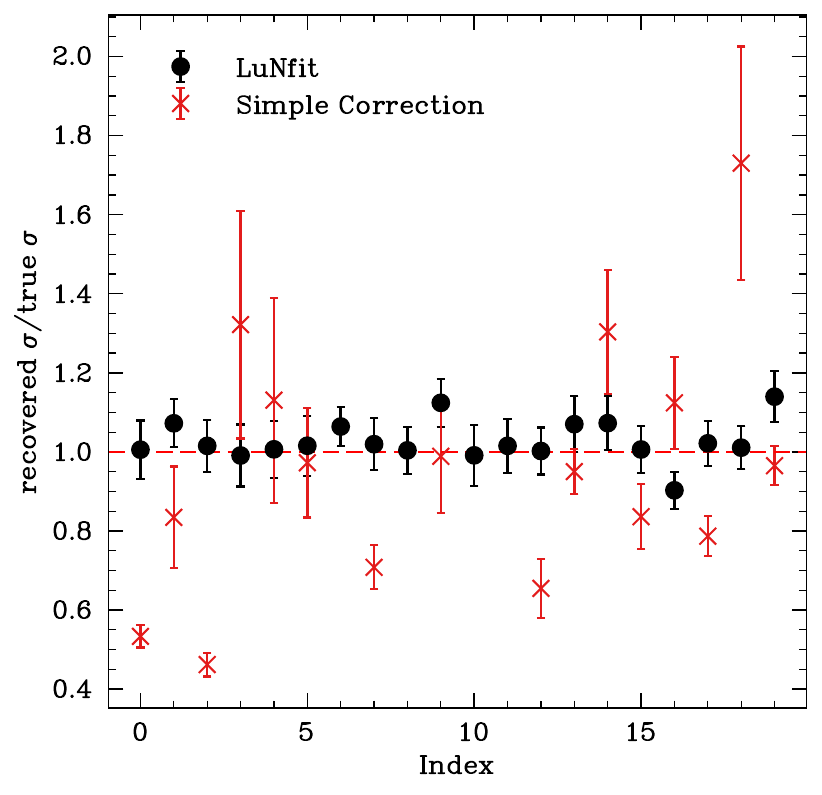}
    \includegraphics[width = .3\textwidth]{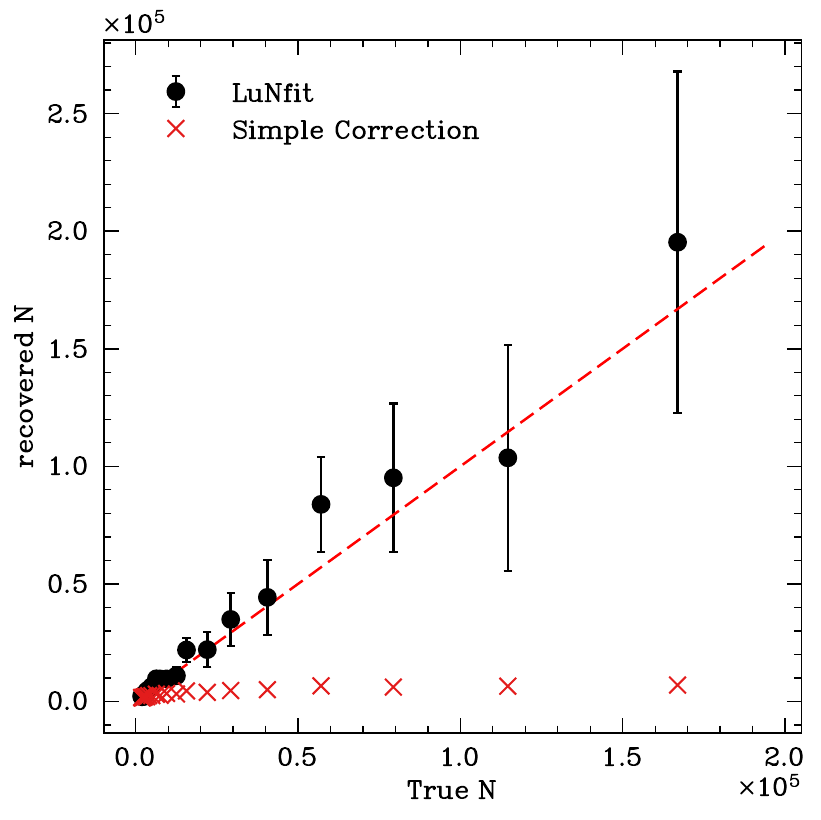}

    \caption{Comparison between a simple correction using the selection function and LuNfit for a log-normal generated distribution with 1000 detections. The red dashed line is the identity line. The "simple correction" is the method described in \cite{10.1038/s41586-021-03878-5}, and a full description is provided in section \ref{sec:comparison}. All simulations are performed with a true $\sigma$ of 0.5.}
    \label{fig:compare-logn}
\end{figure}
The methods in the literature for correcting for the selection effects of single pulses involve estimating the true burst count by scaling up by the lost fraction, e.g., for FRB121102 \citep{10.1038/s41586-021-03878-5}. \red{This method involves sampling the selection effects like LuNfit. Then, they place each detection into a histogram bin. The histogram bin is divided by the detection fraction to recover the ``true count''.} While this method can be effective when the telescope's sensitivity accurately captures the ``turnover'' of the luminosity function, it becomes unreliable in cases where the source is dim, the telescope lacks sensitivity or a combination of both.

Three factors contribute to the limitations of the \cite{10.1038/s41586-021-03878-5} approach. Firstly, accurately measuring the selection effects becomes extremely challenging when the detection fraction is low. As the detection fraction decreases, probing the selection function requires exponentially more computing resources. To address this issue in our analysis, we cut off our selection function at a minimum value of $S_{det}=2$. Secondly, even if the selection effects are well measured, in parameter spaces where the detection fraction is effectively zero, the simple correction used in conventional methods leads to nonsensical results, as the corrected values go to infinity. Lastly, the \cite{10.1038/s41586-021-03878-5} correction requires binning of the detected brightnesses, losing information in the process.

To illustrate the performance of LuNfit compared to the \cite{10.1038/s41586-021-03878-5} method outlined, we provide a comparison in Figure \ref{fig:compare-exp} and \ref{fig:compare-logn}. For the \cite{10.1038/s41586-021-03878-5} method, we correct each histogram bin by dividing by the detection fraction of that bin. We perform a maximum likelihood fit to the scaled bin heights to get the luminosity function parameters. We find that LuNfit exhibits much greater accuracy and precision compared to conventional methods in all cases, especially when the true N is much higher than the detected n.
In summary, LuNfit offers an improved approach to handling selection effects accurately and can provide more reliable results, especially in cases where conventional methods may struggle due to low detection fractions. 

\subsection{Use for intermittent pulsars and comparisons with other tools}
LuNfit has the potential to measure the nulling fraction of many intermittent pulsars and RRATs using the vast observation capabilities of CHIME. While the sample analyzed here is too small to say anything about the population, it does show that there can exist a large range of nulling behavior between 0 and 1 nulling fractions. If raw filterbank data \red{were} saved, then the Gaussian mixture model \citep{10.3847/1538-4357/aaab62} or the \cite{10.1093/mnras/176.2.249} method also can provide comparable results to LuNfit for measurements of the nulling fraction. However, LuNfit has four benefits over the aforementioned tools. Firstly, LuNfit can provide the single pulse luminosity function of intermittent pulsars. Although, arguably, the Gaussian mixture model can also provide a similar measure, this is not possible with the method from \cite{10.1093/mnras/176.2.249}. Secondly, as transient astronomy enters an era of big data, many observatories and programs such as CHIME \citep{10.3847/1538-4357/aad188}, UTMOST \citep{10.1093/mnras/stz1748}, CRAFT \citep{10.1071/AS09082}, MeerTRAP \citep{10.1093/mnras/stac1450} are not saving all observations, rather, only the segments that contain identified pulses by their real-time pipelines. This poses a problem for the Gaussian mixture model and the \cite{10.1093/mnras/176.2.249} method as they require data measurements during non-detections. LuNfit avoids this problem by only requiring a measurement of the selection function, which has already been implemented in some experiments like CHIME/FRB \citep{10.3847/1538-3881/ac9ab5}. Thirdly, the Gaussian mixture method and the \cite{10.1093/mnras/176.2.249} method require identifying the pulse phase where the pulsar emits. If the source is extremely intermittent, where there are only a few single pulses per observation, a few pulses spread across many observations (i.e., some observations contain no detectable emission), or the source is very faint such that even when folded, no appreciable emission is seen, the analysis becomes near impossible with both the Gaussian mixture method and the \cite{10.1093/mnras/176.2.249} method. This is not a problem for LuNfit as long as enough single pulse detections were made on the higher luminosity tail of the distribution. This is because LuNfit does not require pulse phase information, only whether or not detections were made. Finally, LuNfit gives the ability to select between different physically driven models. In this study, we focused on the exponential and log-normal distributions. However, other models, such as the double Gaussian, can be used. LuNfit is based on nested sampling, so the Bayes ratio can always be used to select the preferred model.

\subsection{Potential for repeating FRB studies and limitations}\label{sec:limitations}
We have outlined here the application of LuNfit for RRATs and intermittent slow pulsars. LuNfit can also constrain other radio transients, such as repeating FRBs. Tools like the Gaussian mixture model \citep{10.3847/1538-4357/aaab62} and the \cite{10.1093/mnras/176.2.249} method can not be used on repeating FRBs due to the lack of strict periodicity. In our current implementation of LuNfit, we do not include the width of the pulses in the analysis. This decision is based on the observation that for slow pulsars, the width variations are generally small enough to ignore their impact on the selection function. However, this assumption may not hold for repeating FRBs that exhibit drastic width variability \citep{10.3847/1538-4357/ac33ac}. Therefore, LuNfit can only be applied to repeating FRBs that show minimal width variation. Future iterations of LuNfit will incorporate width dependence to address this issue. As more parameters are introduced, the computational complexity of the analysis increases. Dealing with selection effects in multiple parameters becomes more resource-intensive. Therefore, optimizations will be required to retrieve selection effects while including width variations effectively.\\
\section{Conclusion and future work}
In this study, we demonstrate the importance of accounting for selection biases inherent to the telescope and detection pipeline. This allows the accurate determination of the intrinsic pulse rate and luminosity function of a slow pulsar or RRAT. To achieve this, we created an analysis framework named LuNfit utilizing Bayesian nested sampling, leveraging the \texttt{dynesty} package.

We present detailed simulations and validation procedures for LuNfit in section \ref{sec:methods}. \red{This involved many sets of simulations of fake pulsars and characterizing the nulling fraction of two pulsars, B1905+39 and J2044+4614, where the nulling fraction is known.} As a conclusive application of LuNfit, we apply it to three known RRATs, successfully ascertaining their intrinsic luminosity functions and burst rates. We show that the LuNfit nulling fractions for J1538+2345 align most closely with the FAST observations. We argue that this is likely due to the impressive sensitivity of FAST, enabling them to probe the intrinsic properties of J1538+2345 without much selection bias. The nulling fraction measured by FAST is likely biased slightly high due to short observations. Notably, our findings in section \ref{sec:comparison} highlight the limitations of conventional techniques in capturing the intrinsic luminosity distribution and burst rate \red{when compared with LuNfit}.

Looking ahead, future work includes improving LuNfit by incorporating width dependence and more model types, such as the single and double Gaussian. This extension will render LuNfit more versatile, enabling its effective utilization for sources emitting bursts with complex and varying morphologies. This enhancement will allow LuNfit to be applied to all repeating Fast Radio Bursts (FRBs), pulsars, and radio magnetars. Additionally, harnessing LuNfit's current capabilities, we intend to quantify the nulling fraction for numerous intermittent pulsars and RRATs using CHIME/Pulsar observations. This will provide a systematic understanding of the nulling slow pulsar population. \red{Finally, we plan to make the codebase publicly available with the next iteration of LuNfit in the near future.}

\section{Acknowledgements}

We acknowledge that CHIME is located on the traditional, ancestral, and unceded territory of the Syilx/Okanagan people. We are grateful to the staff of the Dominion Radio Astrophysical Observatory, which is operated by the National Research Council of Canada.  CHIME is funded by a grant from the Canada Foundation for Innovation (CFI) 2012 Leading Edge Fund (Project 31170) and by contributions from the provinces of British Columbia, Qu\'{e}bec and Ontario. The CHIME/FRB Project, which enabled development in common with the CHIME/Pulsar instrument, is funded by a grant from the CFI 2015 Innovation Fund (Project 33213) and by contributions from the provinces of British Columbia and Qu\'{e}bec, and by the Dunlap Institute for Astronomy and Astrophysics at the University of Toronto. Additional support was provided by the Canadian Institute for Advanced Research (CIFAR), McGill University and the McGill Space Institute thanks to the Trottier Family Foundation, and the University of British Columbia. The CHIME/Pulsar instrument hardware was funded by NSERC RTI-1 grant EQPEQ 458893-2014. This research was enabled in part by support provided by the BC Digital Research Infrastructure Group and the Digital Research Alliance of Canada (alliancecan.ca).

\noindent F.A.D is supported by the UBC Four Year Fellowship

 \noindent D.C.S. is supported by an NSERC Discovery Grant (RGPIN-2021-03985) and by a Canadian Statistical Sciences Institute (CANSSI) Collaborative Research Team Grant.

 \noindent RVC is supported by an NSERC Discovery Grant  RGPIN-2018-05663

 \noindent A.B.P. is a Banting Fellow, a McGill Space Institute~(MSI) Fellow, and a Fonds de Recherche du Quebec -- Nature et Technologies~(FRQNT) postdoctoral fellow.

 \noindent GME is supported by a CANSSI Collaborative Research Team Grant with support from NSERC, and an NSERC Discovery Grant RGPIN-2020-04554.

 \noindent Pulsar and FRB research at UBC are supported by an NSERC Discovery Grant and by the Canadian Institute for Advanced Research\\

%

\vspace{5mm}
\facilities{CHIME}


\software{PRESTO \citep{2001PhDT.......123R}, CHIPSPIPE \citep{10.1093/mnras/stad2012}, Numpy \citep{10.1038/s41586-020-2649-2}, Scipy \citep{10.1038/s41592-019-0686-2}, sigpyproc3}

\appendix

\section{Simulations}
\begin{figure}[ht]
\centering
\gridline{
          \fig{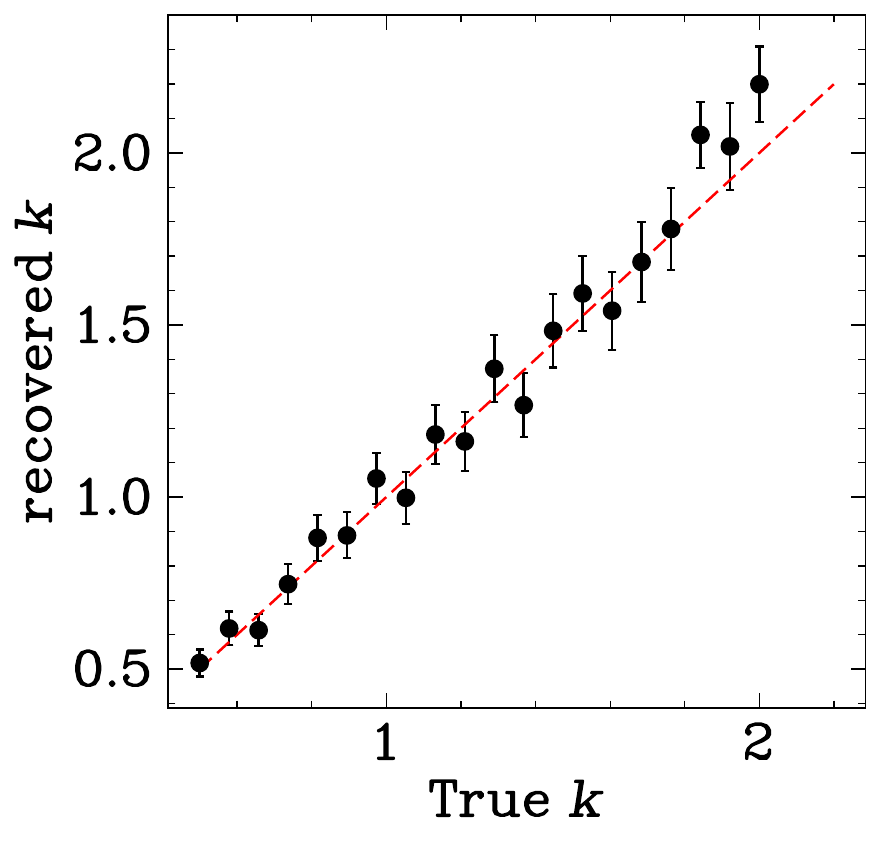}{0.24\textwidth}{(a) 150 detections}
  \fig{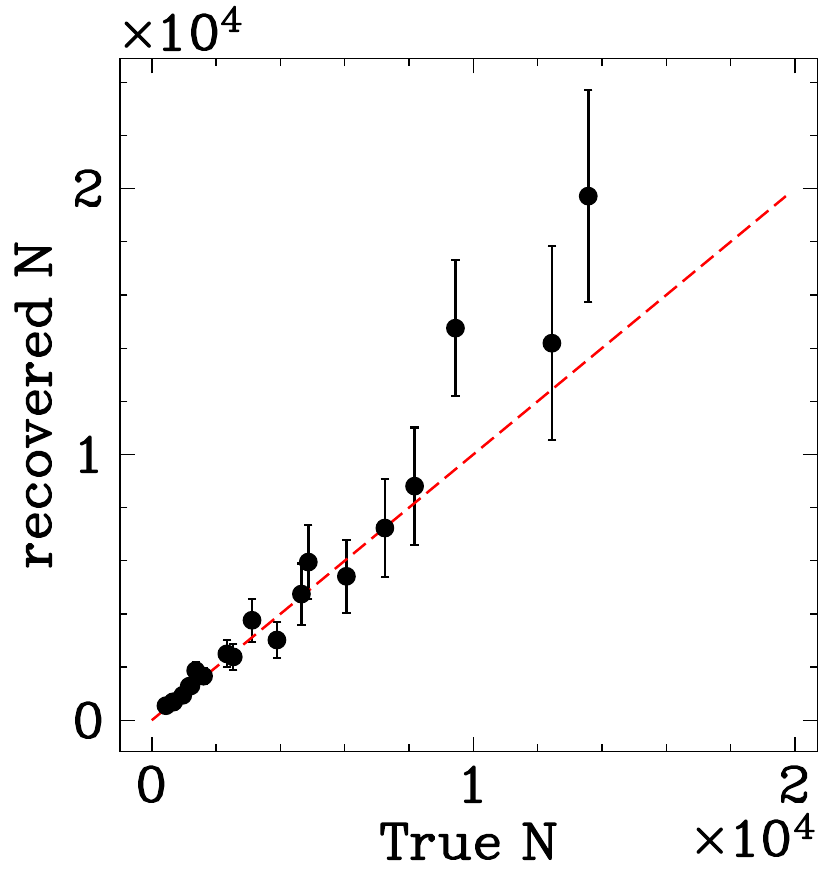}{0.24\textwidth}{(b) 150 detections}
            \fig{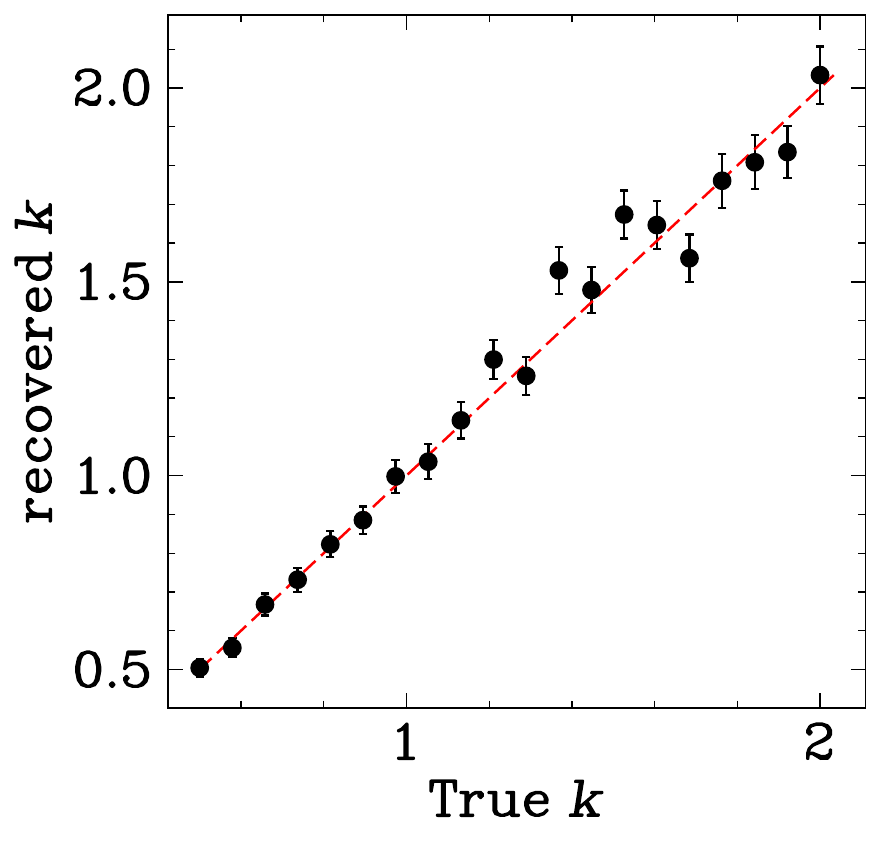}{0.24\textwidth}{(c) 500 detections}
            \fig{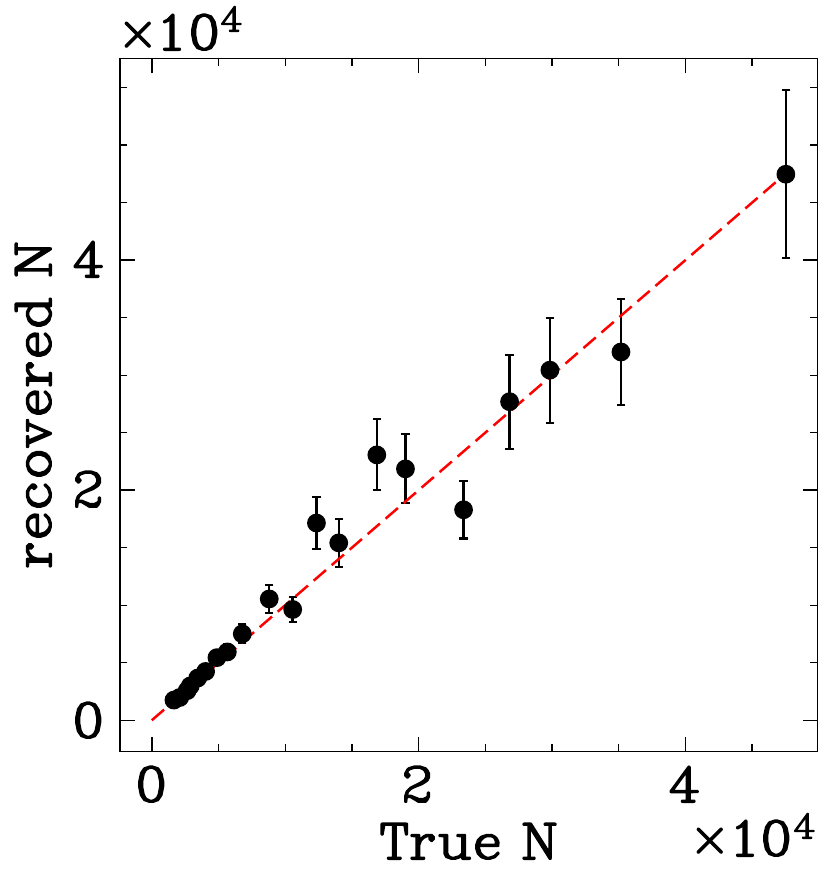}{0.2\textwidth}{(d) 500 detections}
          }
\gridline{
          \fig{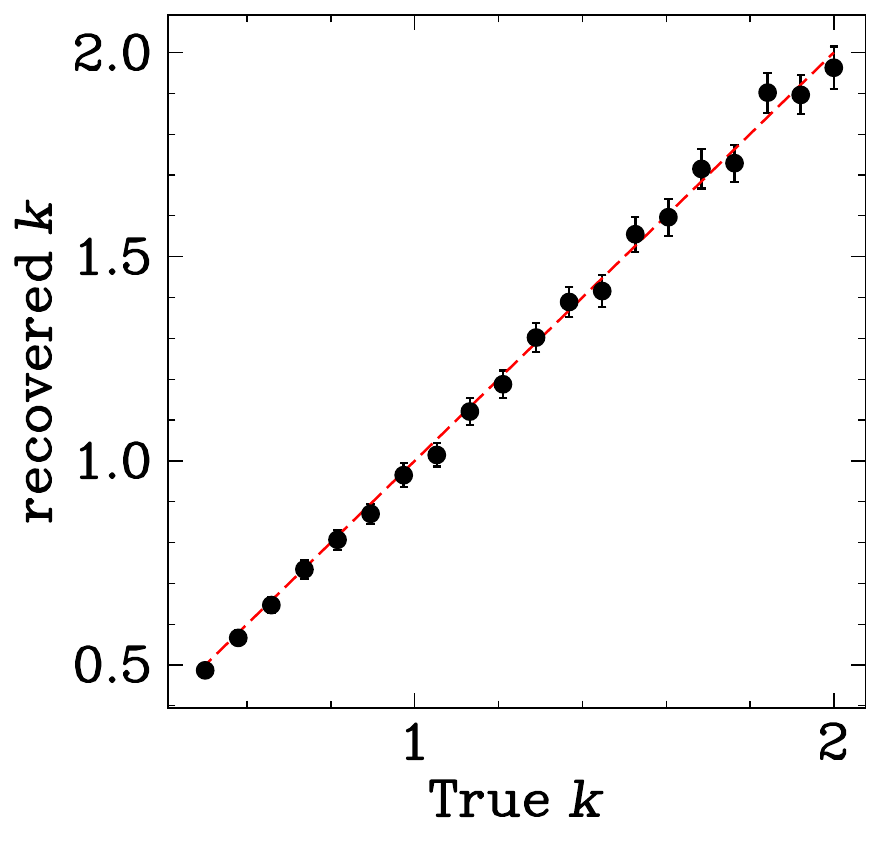}{0.24\textwidth}{(e) 1000 detections}
        \fig{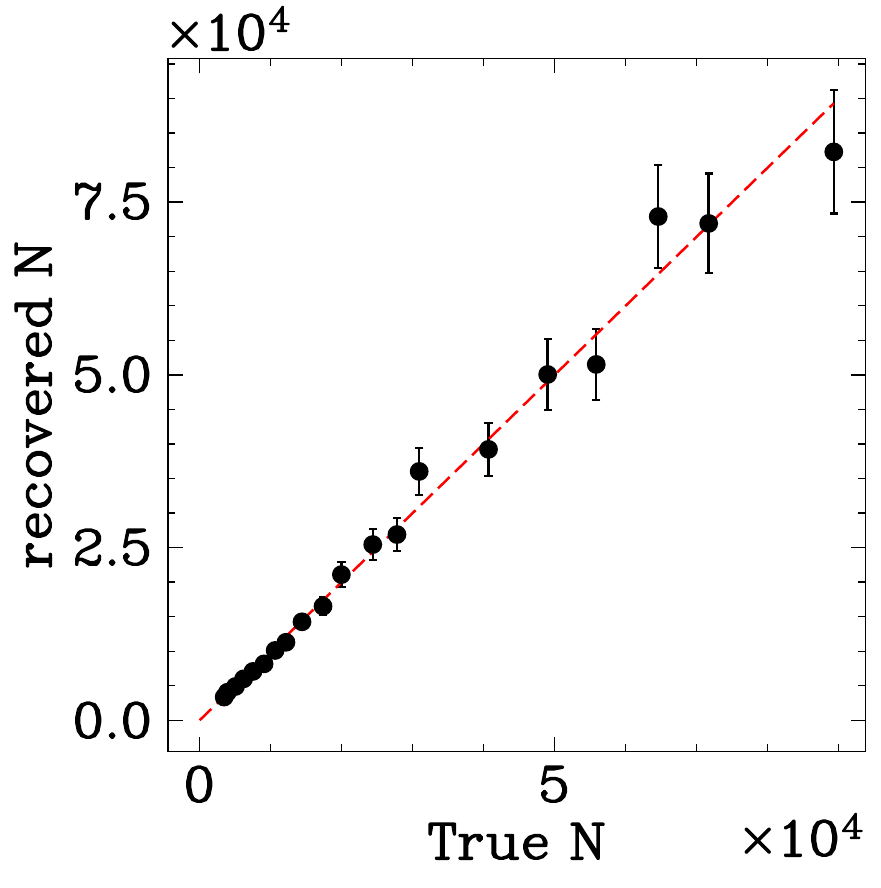}{0.24\textwidth}{(f) 1000 detections}
        \fig{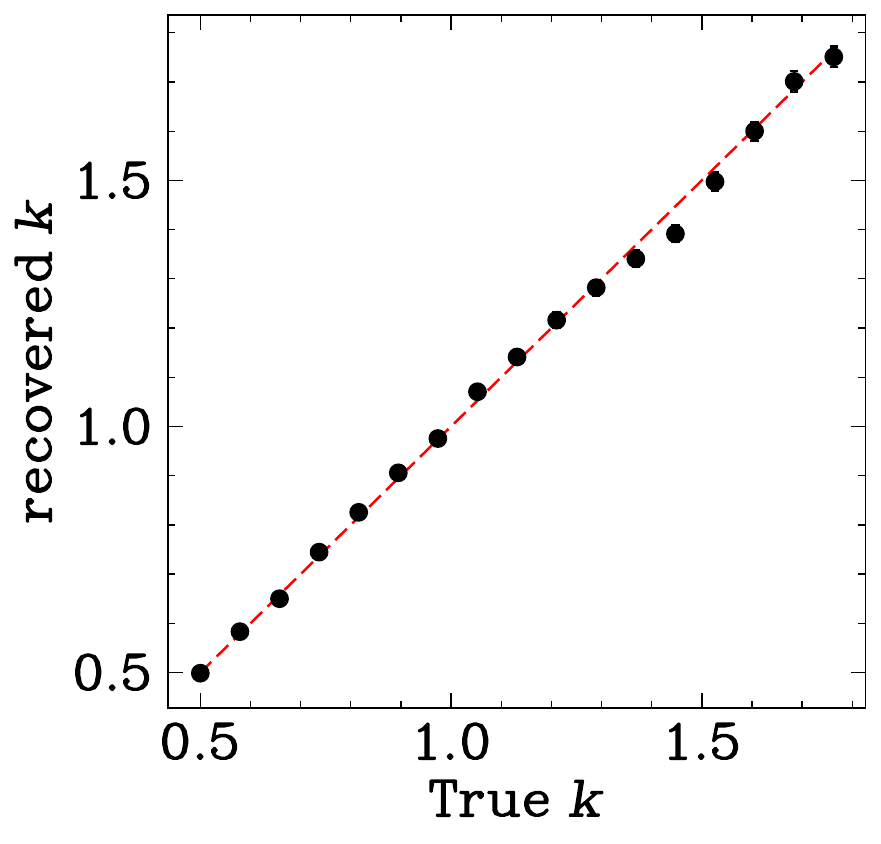}{0.24\textwidth}{(g) 5000 detections}
          \fig{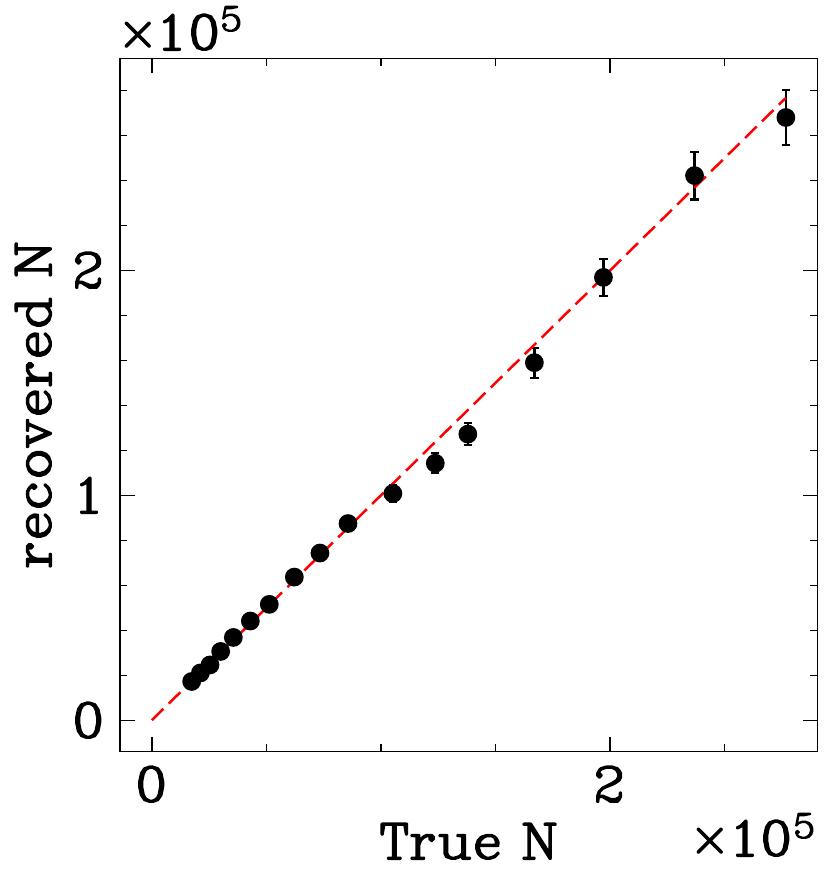}{0.2\textwidth}{(h) 5000 detections}
          }
          \caption{Simulations and recovery capabilities of LuNfit for the exponential distribution. \red{The axes show the LuNfit recovered quantities against the true simulated quantities}. N and k are the parameters constrained for an exponential luminosity distribution. \red{The red dashed line is the identify line. All points should lie on this line for a perfect fitting algorithm. This figure shows the recovered k (N) value given a simulated (true) k (N) value. As expected for high true N, the error bars increase as this corresponds to an intrinsically fainter pulsar. In other words, LuNfit needs to predict more missed pulses for higher true N values. For an exponential distribution, a fainter pulsar corresponds to a larger k. Therefore, we similarly find that the errors for k increase for larger k. LuNfit performs significantly better for an exponential distribution compared to a log-normal distribution due to the smaller parameter space.}}
          \label{fig:lunfit_exp}
\end{figure}
\begin{figure}[ht]
\centering
\gridline{\fig{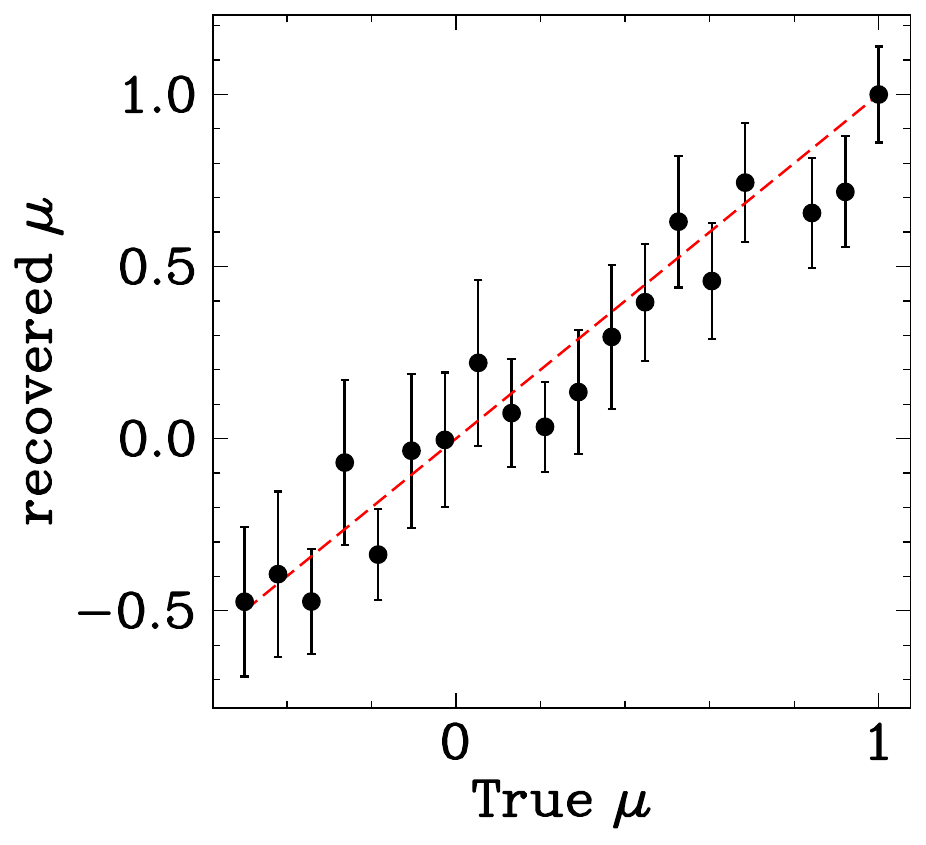}{0.25\textwidth}{(a) 150 detections}
          \fig{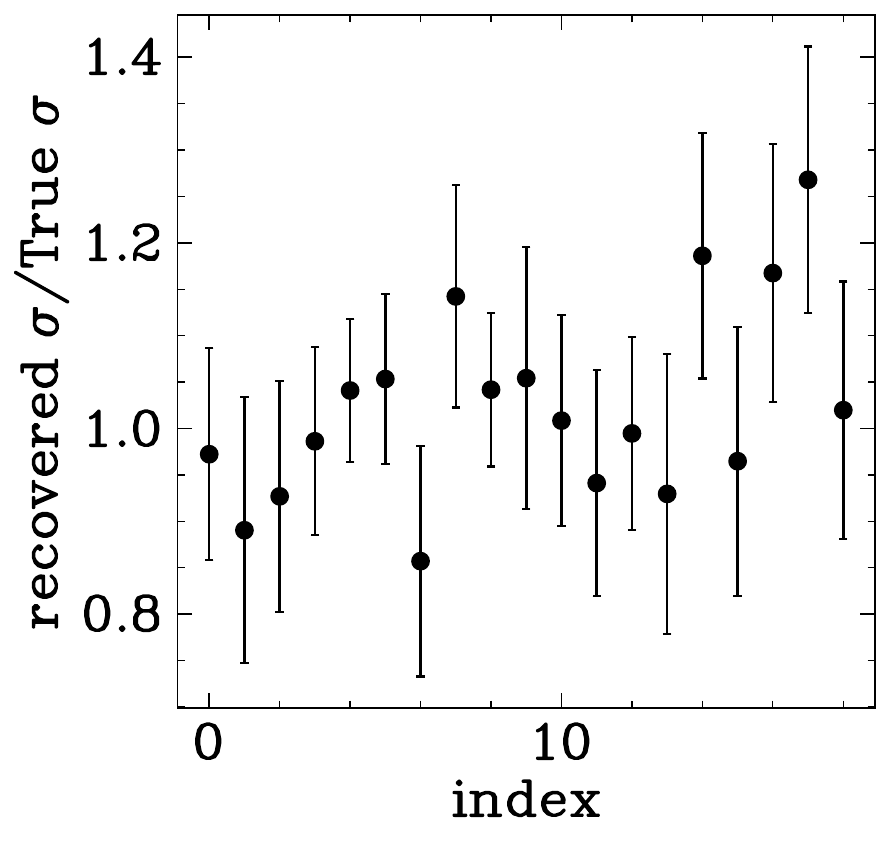}{0.25\textwidth}{(b) 150 detections}
        \fig{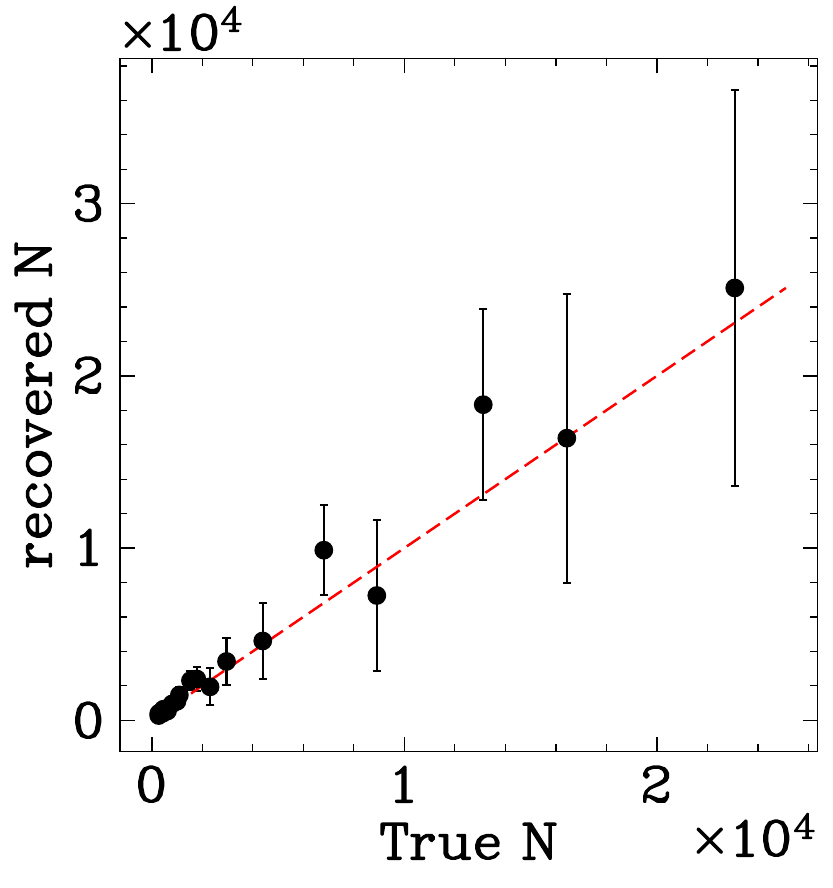}{0.235\textwidth}{(c) 150 detections}
          }

\gridline{\fig{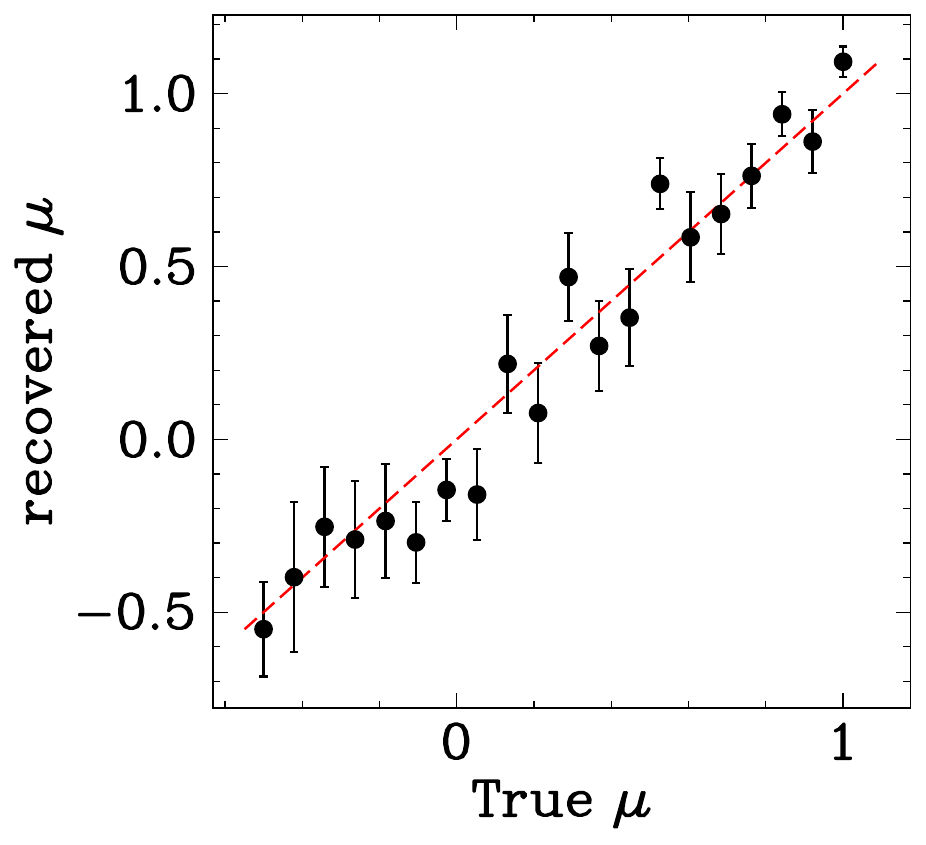}{0.25\textwidth}{(d) 500 detections}
          \fig{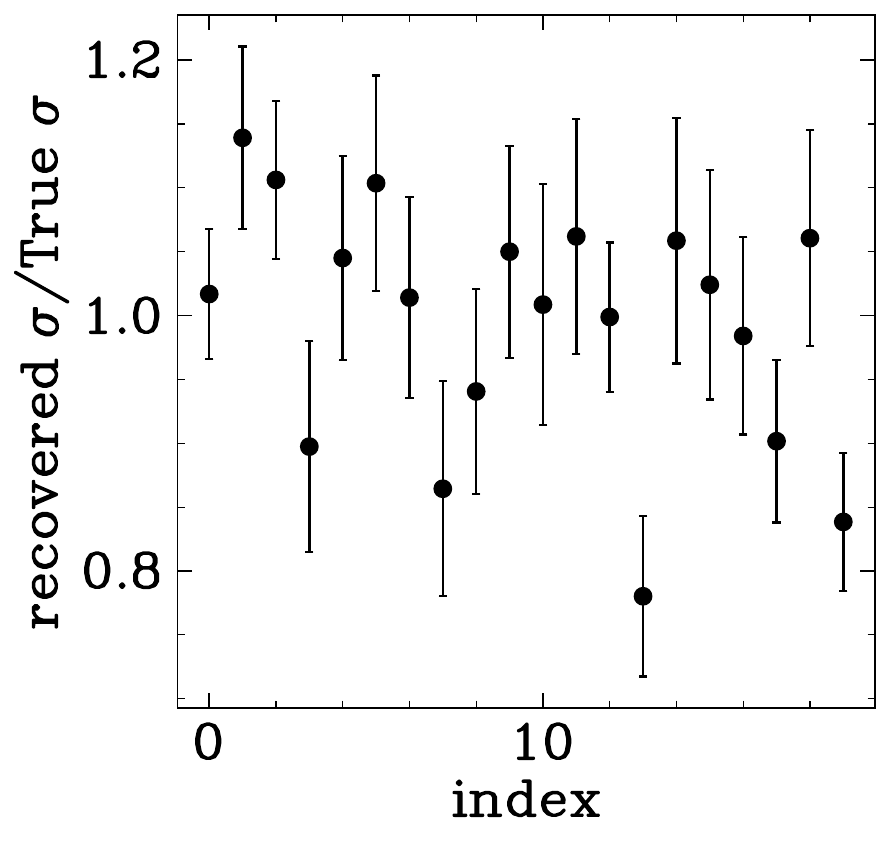}{0.25\textwidth}{(e) 500 detections}
        \fig{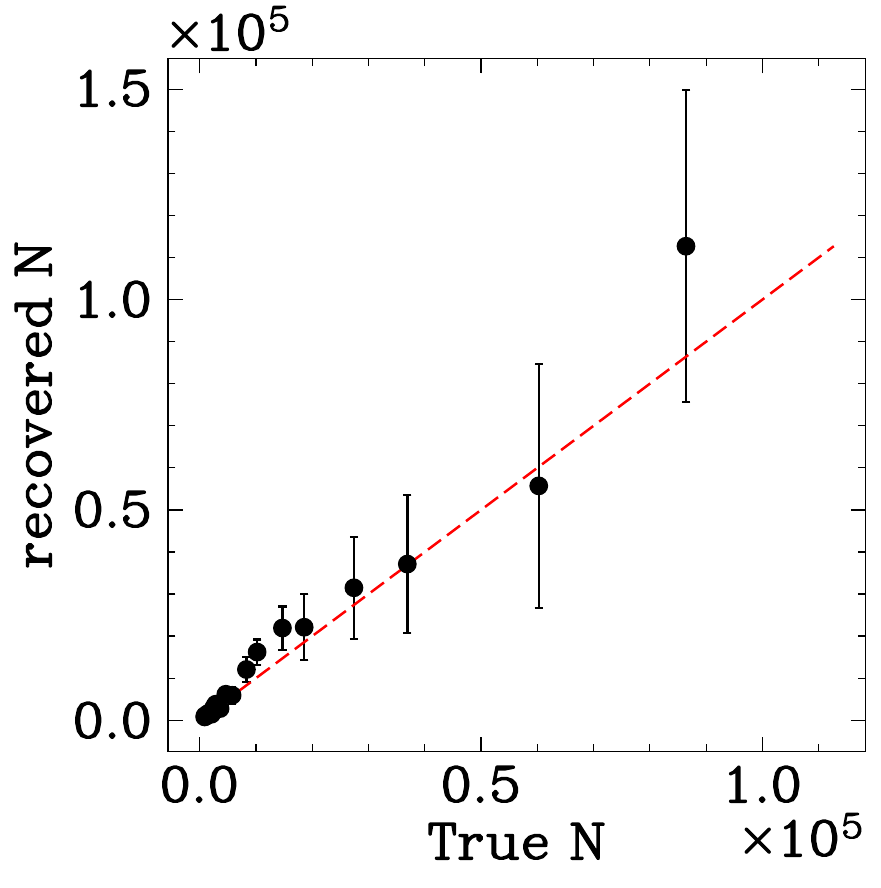}{0.235\textwidth}{(f) 500 detections}
          }
\gridline{\fig{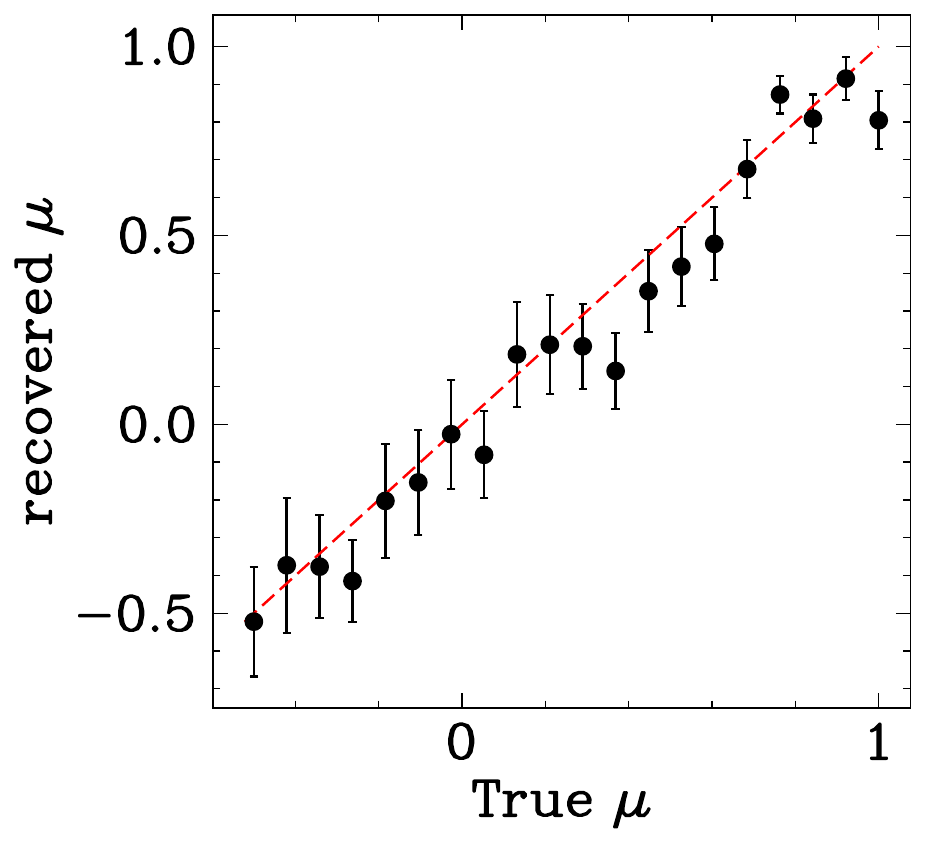}{0.25\textwidth}{(g) 1000 detections}
          \fig{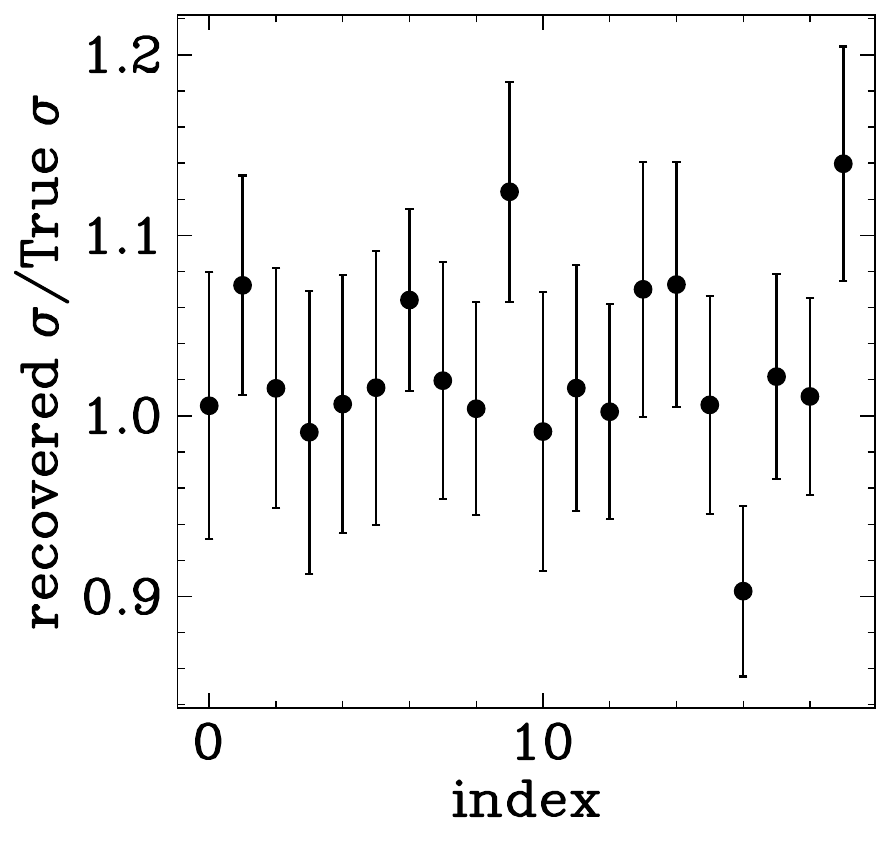}{0.25\textwidth}{(h) 1000 detections}
        \fig{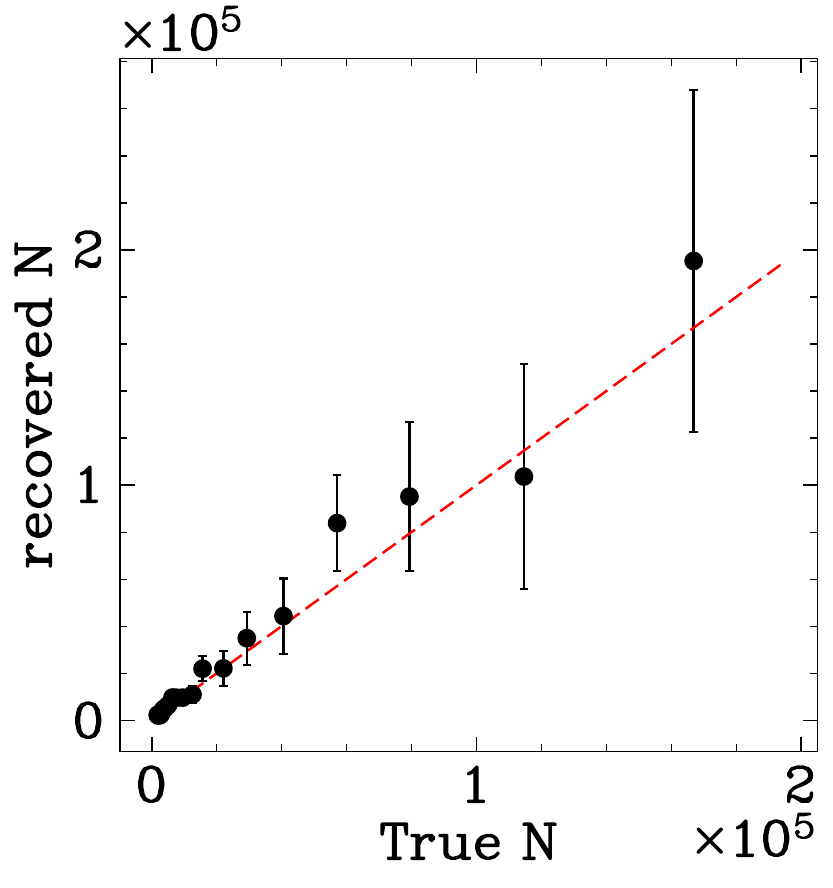}{0.235\textwidth}{(i) 1000 detections}
          }

  \gridline{\fig{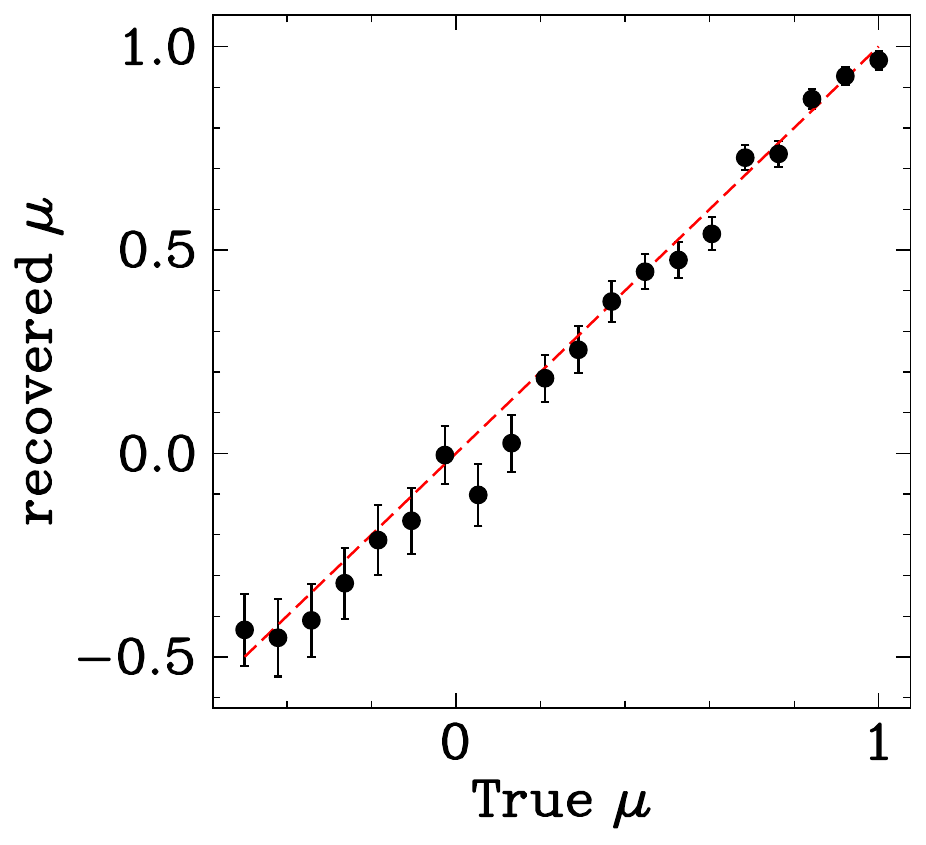}{0.25\textwidth}{(j) 5000 detections}
          \fig{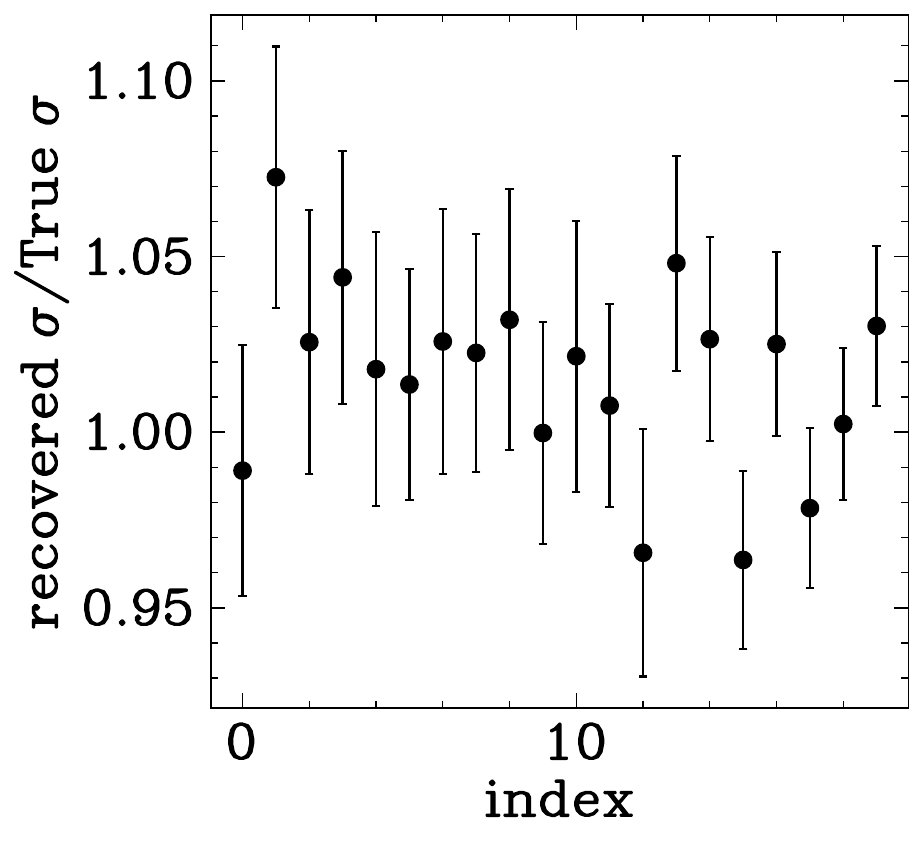}{0.25\textwidth}{(k) 5000 detections}
        \fig{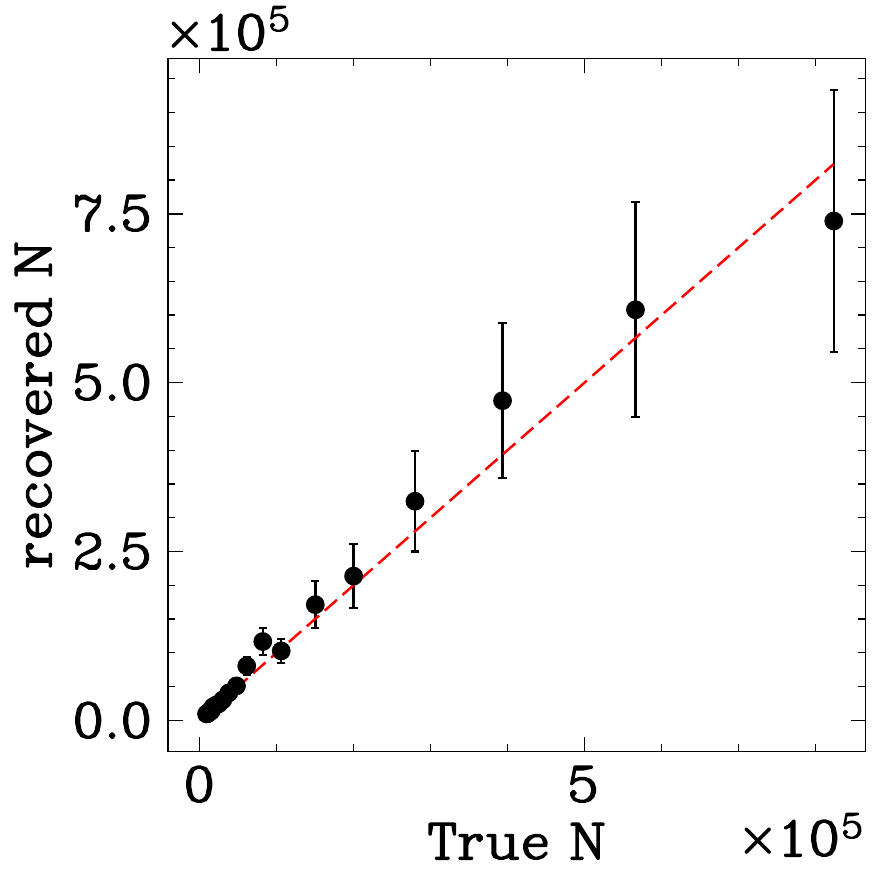}{0.235\textwidth}{(l) 5000 detections}
          }
        \caption{Simulations and recovery capabilities of LuNfit for the log-normal distribution. \red{The axes show the LuNfit recovered quantities against the true simulated quantities}. $\mu$, $\sigma$, and N are the parameters constrained for a log-normal luminosity distribution. All simulations shown here have a $\sigma$ of 0.5. \red{Again, the red dashed line is the identity line. All points should fall on this line for a perfect fitting algorithm. Like the exponential distribution shown in Figure \ref{fig:lunfit_exp}, the fit is worse for high true N and low true $\mu$ as this indicates a fainter pulsar. The uncertainties of the log-normal luminosity model are markedly larger than that of the exponential luminosity model. This is because of the larger parameter space to explore. All simulations were performed with $\sigma=0.5$; thus, all data points in the middle column should equal 1 (within uncertainty). This figure shows that in most cases LuNfit is able to retrieve the correct $\sigma$ value.}}
          \label{fig:lunfit_logn}
\end{figure}
\begin{figure}[ht]
\centering
\gridline{\fig{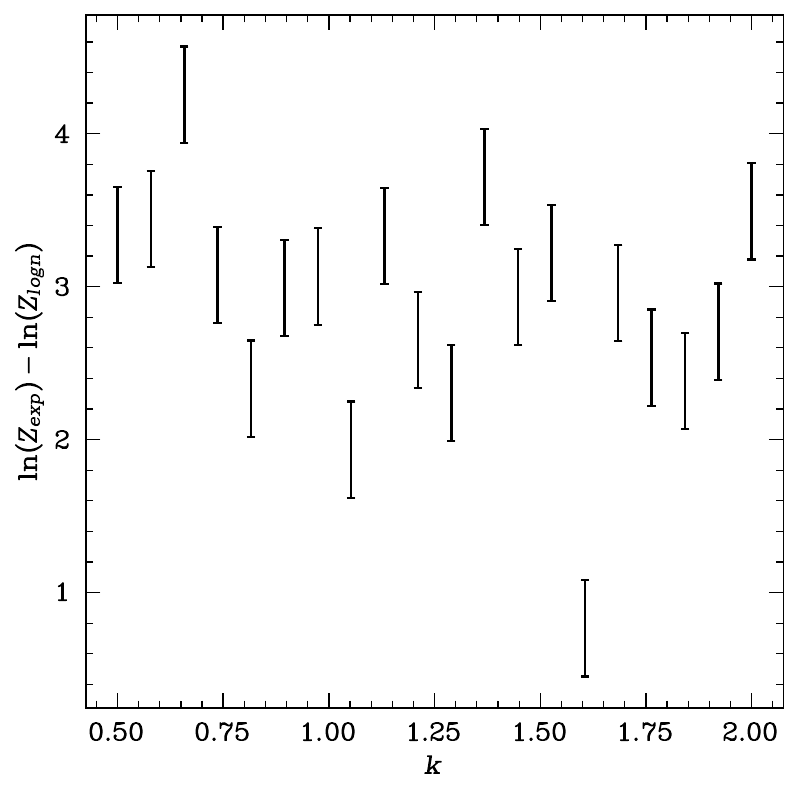}{0.48\textwidth}{(a) 150 detections}
          \fig{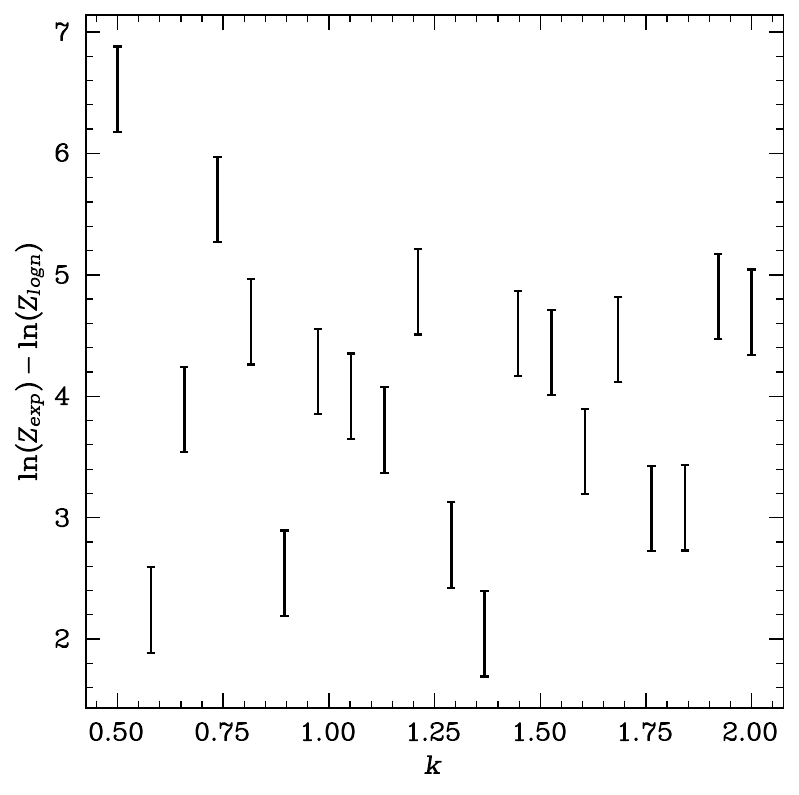}{0.48\textwidth}{(b) 500 detections}
          }
          \gridline{
            \fig{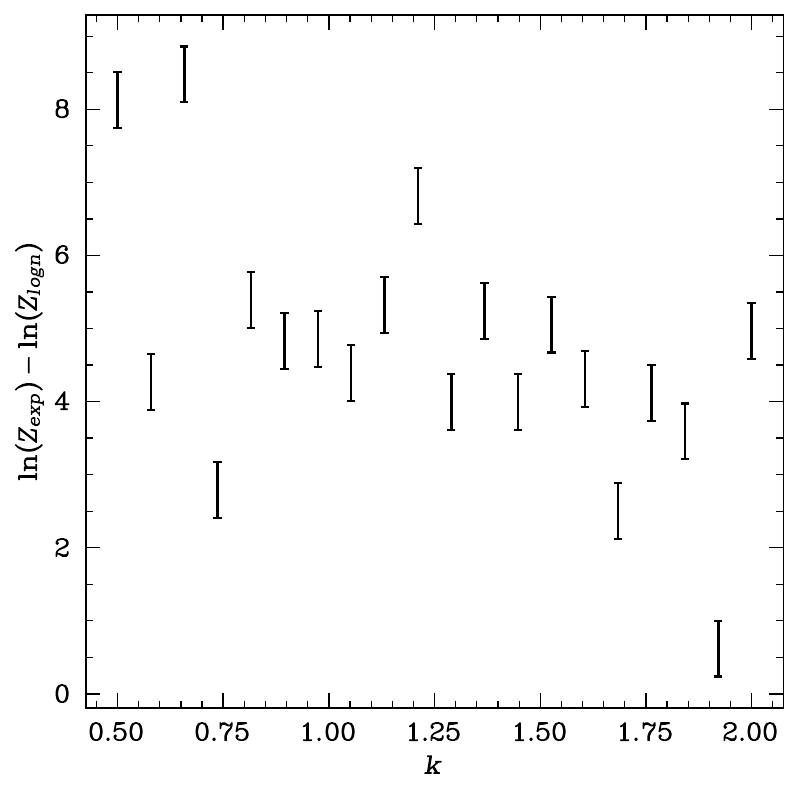}{0.48\textwidth}{(c) 1000 detections}
          \fig{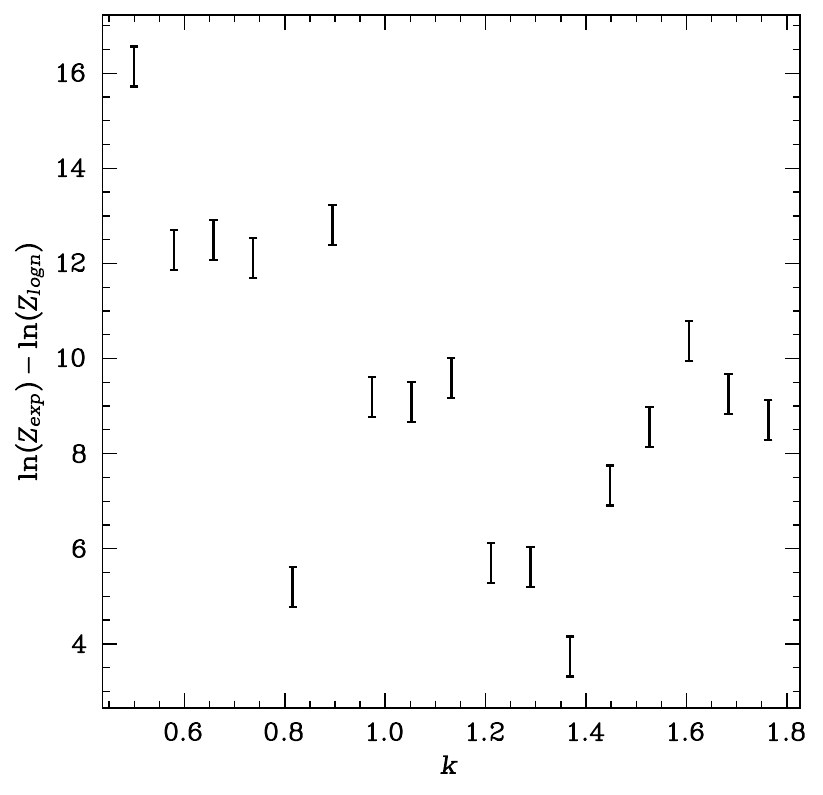}{0.48\textwidth}{(d) 5000 detections}}
          \caption{Bayes ratio for a simulated exponential distribution.  The Bayes ratio is given in favor of the exponential distribution for four sets of exponential luminosities distributions. \red{That is, positive values indicate a preference for the exponential distribution.} There are 150 (a), 500(b), 1000(c), and 5000(d) detected pulses in each simulation. \red{If the true underlying distribution is exponential, LuNfit always correctly identifies this, as shown by the absence of negative values. In general, the Bayes ratio is larger for low k and a larger number of detections.}}
          \label{fig:bayes_exp}
\end{figure}
\begin{figure}[ht]
\centering
\gridline{\fig{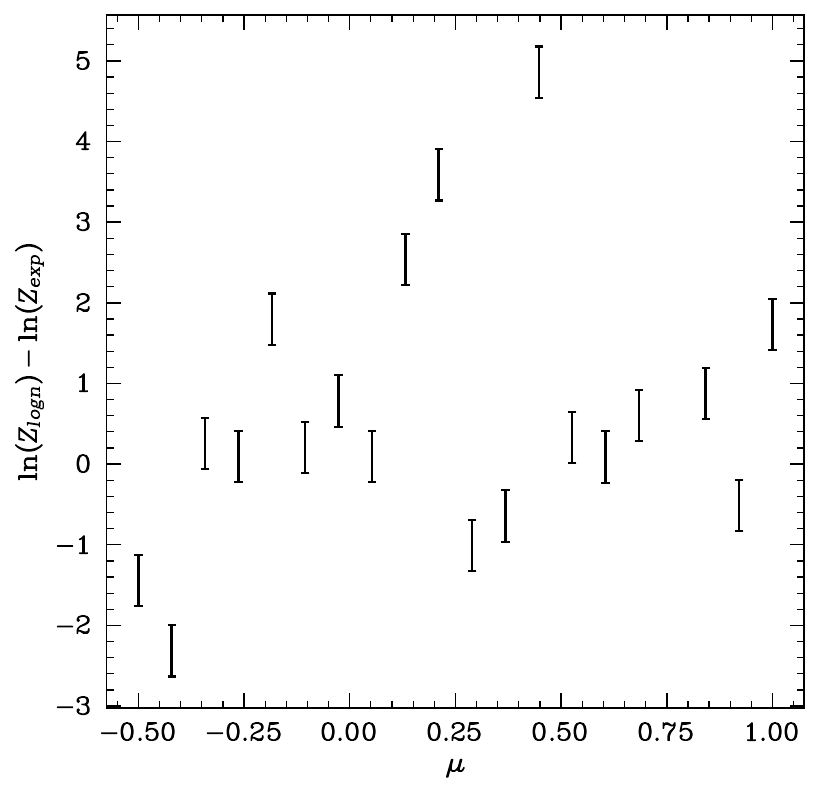}{0.48\textwidth}{(a) 150 detections}
          \fig{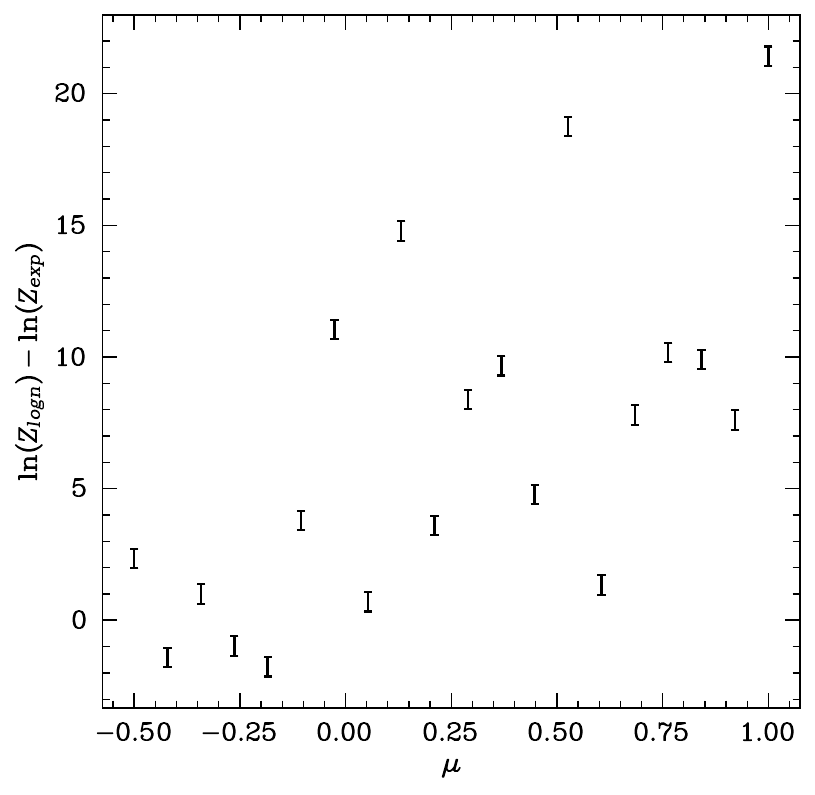}{0.48\textwidth}{(b) 500 detections}
          }
          \gridline{
            \fig{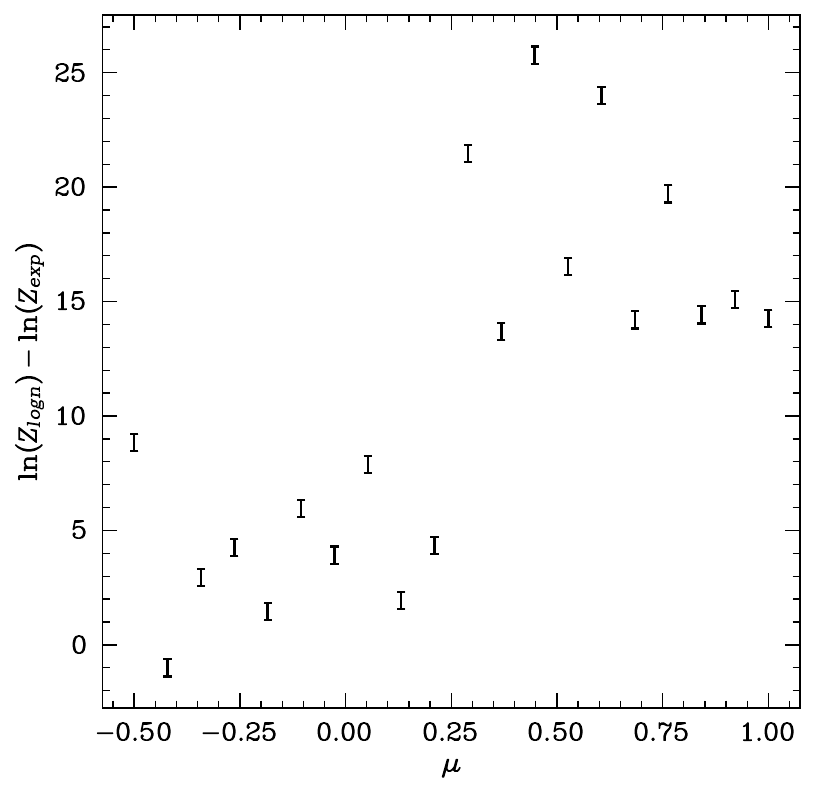}{0.48\textwidth}{(c) 1000 detections}
          \fig{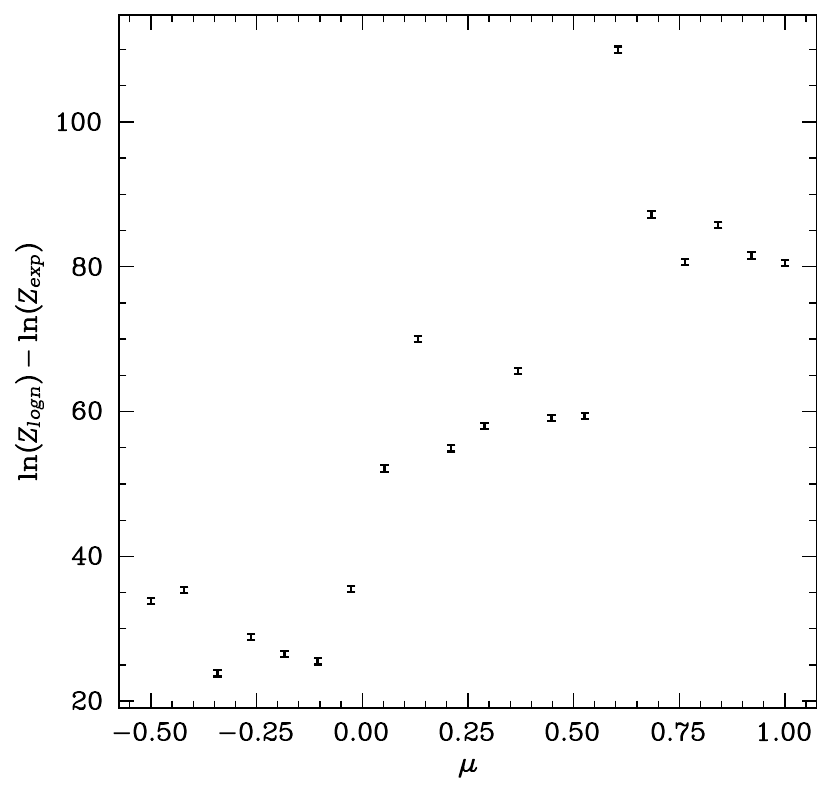}{0.48\textwidth}{(d) 5000 detections}}
          \caption{Bayes ratio for simulated log-normal distributions. The Bayes ratio is given in favor of the log-normal distribution for four sets of log-normal luminosities. \red{That is, positive values indicate a preference for a log-normal distribution.} There are 150 (a), 500(b), 1000(c) and 5000(d) detected pulses in each simulation. The $\sigma$ for these simulations was set to 0.5. \red{The effects of a brighter median luminosity (higher $\mu$) are much more pronounced for the log-normal distribution than the exponential distribution. Higher $\mu$ values are clearly positively correlated with higher Bayes ratios. This is due to LuNfit being able to completely fit the ``turn-over'' in log-normal distributions. Additionally, the log-normal tends towards an exponential for low $\mu$ values. Once again, a larger number of detections aid the differentiation between log-normal and exponential distributions.}}
          \label{fig:bayes_logn}
\end{figure}

\clearpage

\bibliographystyle{aasjournal}
\bibliography{rrat_or_not}{}

\end{document}